\documentclass[12pt]{article}

\usepackage{amssymb,amsmath}
\usepackage{graphicx}
\usepackage{geometry}

\usepackage{natbib}

\usepackage[lined,boxed,ruled]{algorithm2e}

\usepackage[tight]{subfigure}
\usepackage{color}

\definecolor{sky}{rgb}{0.14,0.45, 0.61}
\definecolor{yell}{rgb}{0.66, 0.66, 0.15}
\definecolor{darkgray}{rgb}{0.2, 0.2, 0.2}

\begin{document}

\title{Self-organizing traffic lights\\at multiple-street intersections}
\author{Carlos Gershenson \and David A.\ Rosenblueth\\
Departamento de Ciencias de la Computaci\'on\\
Instituto de Investigaciones en Matem\'aticas Aplicadas y en Sistemas\\
Universidad Nacional Aut\'onoma de M\'exico\\
A.\ P.\ 20-726,
01000 M\'exico D.F.\ M\'exico\\
Tel. +52 55 56 22 36 19 \
Fax +52 55 56 22 36 20 \\
\href{mailto:cgg@unam.mx}{cgg@unam.mx} \
\url{http://turing.iimas.unam.mx/~cgg} \\
\href{mailto:drosenbl@servidor.unam.mx}{drosenbl@servidor.unam.mx} \ \url{http://leibniz.iimas.unam.mx/~drosenbl/}\\
}
\maketitle

\begin{abstract}
The elementary cellular automaton following rule 184 can mimic 
particles flowing in one direction at a constant speed.
This automaton can therefore model highway traffic.
In a recent paper, we have incorporated
intersections regulated by traffic lights to this model using
exclusively elementary cellular automata.
In such a paper, however, we only explored a rectangular grid.
We now extend our model to more complex scenarios employing an
hexagonal grid.
This extension shows first that our model can readily incorporate
multiple-way intersections and hence simulate complex scenarios.
In addition, the current extension allows
us to study and evaluate the behavior of two different kinds of traffic
light controller for a grid of six-way streets allowing for either two
or three street intersections: a traffic
light that 
tries to \emph{adapt} to the amount of traffic (which results in 
self-organizing traffic lights) and a system of synchronized traffic
lights with coordinated
rigid periods (sometimes called the ``green wave'' method).
We observe a tradeoff between system capacity and topological
complexity.
The green wave method is unable to cope with the complexity of a
higher-capacity scenario, while the self-organizing method is
scalable, adapting to the complexity of a scenario and exploiting its
maximum capacity.
Additionally, in this paper we propose a  benchmark, independent of methods and models, to measure the performance of a traffic light controller comparing it against a theoretical optimum.

%

\begin{center}
{\bf Nontechnical Abstract}
\end{center}

Traffic light coordination is a complex problem. In this paper, we extend previous work on an abstract model of city traffic to allow for multiple street intersections. We test a self-organizing method in our model, showing that it is close to theoretical optima and superior to a traditional method of traffic light coordination.

\end{abstract}


\section{Introduction}
The purpose of any model is to simplify reasoning while illuminating
reality.
Elementary cellular automata ~\citep{Wolfram1986,WuenscheLesser1992,Wolfram:2002} are especially interesting models
because of being at the simplicity end of the spectrum, and for at the
same time exhibiting complex behavior.
The elementary cellular automaton following rule 184, for example, can
simulate particles moving in one direction at a constant speed.
This property suggests using such an automaton for modeling vehicular traffic.
Indeed, there have been several traffic models based on cellular automata
(CA)~\citep{cremer:ludwig:86,NaSch1992,biham:middleton:levine:92,nagel:paczuski:95,fukui:ishibashi:96,chopard:luthi:queloz:96,simon:nagel:97,chowdhury:schadschneider:99,schadschneider&:99,BrockfeldEtAl2001}. 
We are interested in modeling intersections with traffic lights so as
to compare different methods of controlling such intersections.
In a previous work~\citep{RosenbluethGershenson:2010} we have shown how to model traffic-light
intersections with elementary cellular automata following, in addition
to rule 184, different rules depending on whether or not the light is
allowing vehicles to flow.
In that work, we only considered rectangular grids.
Our purpose will be, first, to show that elementary cellular automata
can also model hexagonal grids (allowing the possibility of incorporating
3-way intersections), and second, to evaluate different traffic-light systems.
Our results show both the scalability of our approach and the
capacity of our method to illuminate the relationships between street
topology, intersection capacity, and traffic controllers.





We evaluate two different methods of traffic light coordination: a traditional ``\emph{green wave}" method that tries to optimize phases according to expected traffic flows, and a previously proposed \emph{self-organizing} method \citep{Gershenson2005,CoolsEtAl2007,GershensonRosenblueth:2010}
 extended to a more complex scenario.
We also compare results with theoretical optimality curves that our CA model of city traffic can provide straighforwardly and with a random assignment of phases as a worst-case scenario.
It is to be expected that the adaptable, \emph{self-organizing} method will
be superior to the fixed-period, \emph{green wave} method.
A reason is that, unlike the fixed-period lights, the self-organizing
lights have sensors.
These sensors provide feedback which should pay off resulting in a
better traffic-light method.
What is interesting is that, unlike the \emph{green wave} method, which has
a central control, the \emph{self-organizing} method has no such control:
The traffic lights communicate with each other using as signals the vehicles
traveling from one intersection to the next.

The simulations revealed, as expected, the superiority of the
\emph{self-organizing} method over the \emph{green wave} approach, approaching or matching the optimality curves for a broad range of densities. 
Moreover, the simulations also showed a number of interesting
characteristics of the self-organizing method, enabling us to 
discover up to 10 phase transitions and close-to-optimal performance.

In the next section, we review an extension of the rule 184 model of highway traffic where intersections can be incorporated, with the possibility of modeling city traffic at a very abstract level. We extend our previous work to compare traffic light controllers with theoretical optima.
In section \ref{sec:methods}, the methods for coordinating traffic lights are introduced. Section \ref{sec:single} presents results of simple scenarios with only three streets, while section \ref{sec:city} shows results of more complex scenarios. Discussion follows in section \ref{sec:discussion} and conclusions in section \ref{sec:conclusions}. 
The \emph{self-organizing} method is detailed in Appendix \ref{app:sola}.


\section{An elementary model of city traffic}
\label{sec:model}

In CA models, time and space are discrete. 
Vehicles are represented within cells, and rules determine how and whether vehicles ``move''.
The simplest highway traffic model proposed is elementary
cellular automaton ``Rule 184"~\citep{Yukawa:1994,ChowdhuryEtAl2000,Maerivoet:2005}.

\begin{figure}
     \centering
     \subfigure[]{
          \label{fig:184A}
          \includegraphics[width=.45\textwidth]{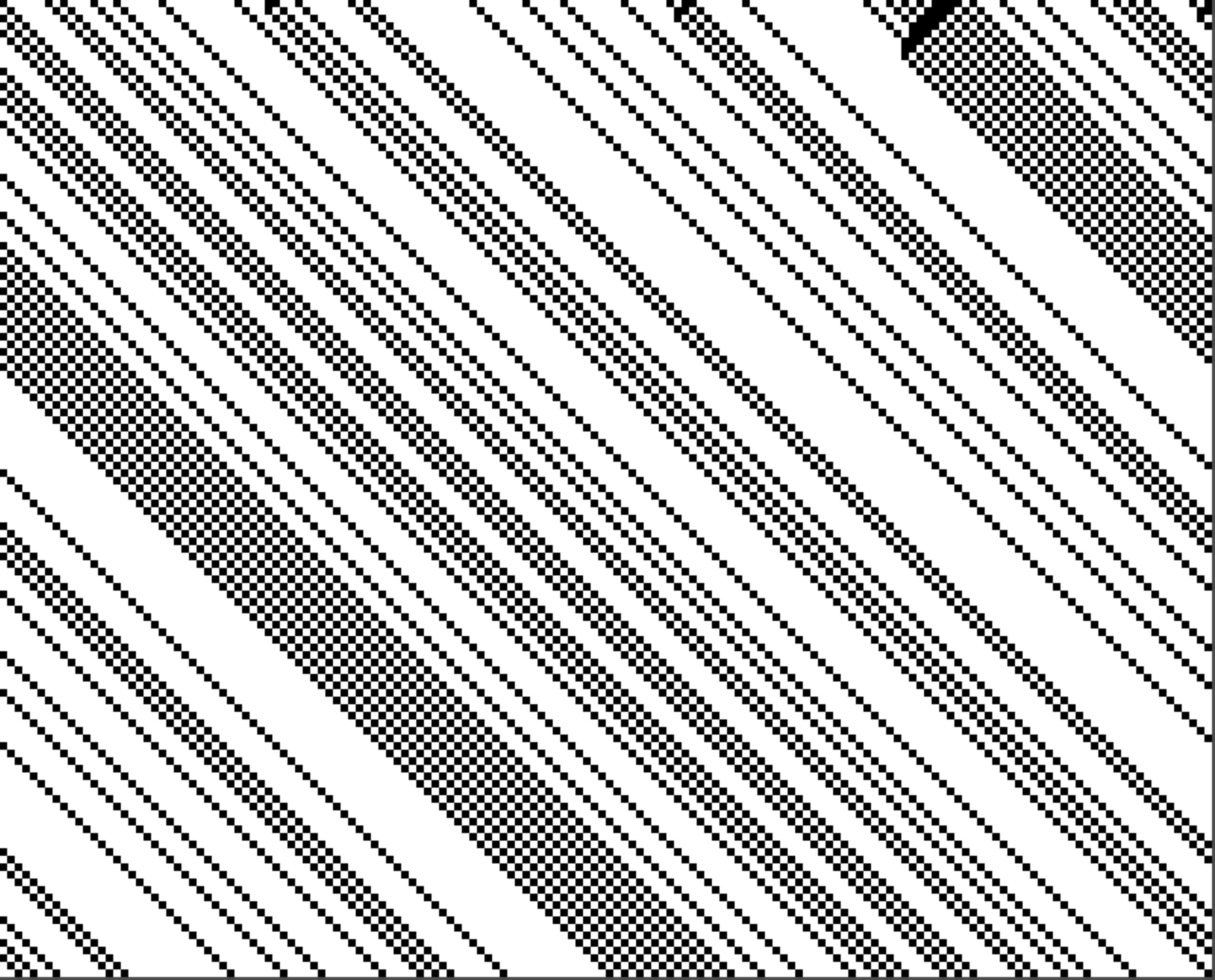}
	}
     \subfigure[]{
          \label{fig:184B}
          \includegraphics[width=.45\textwidth]{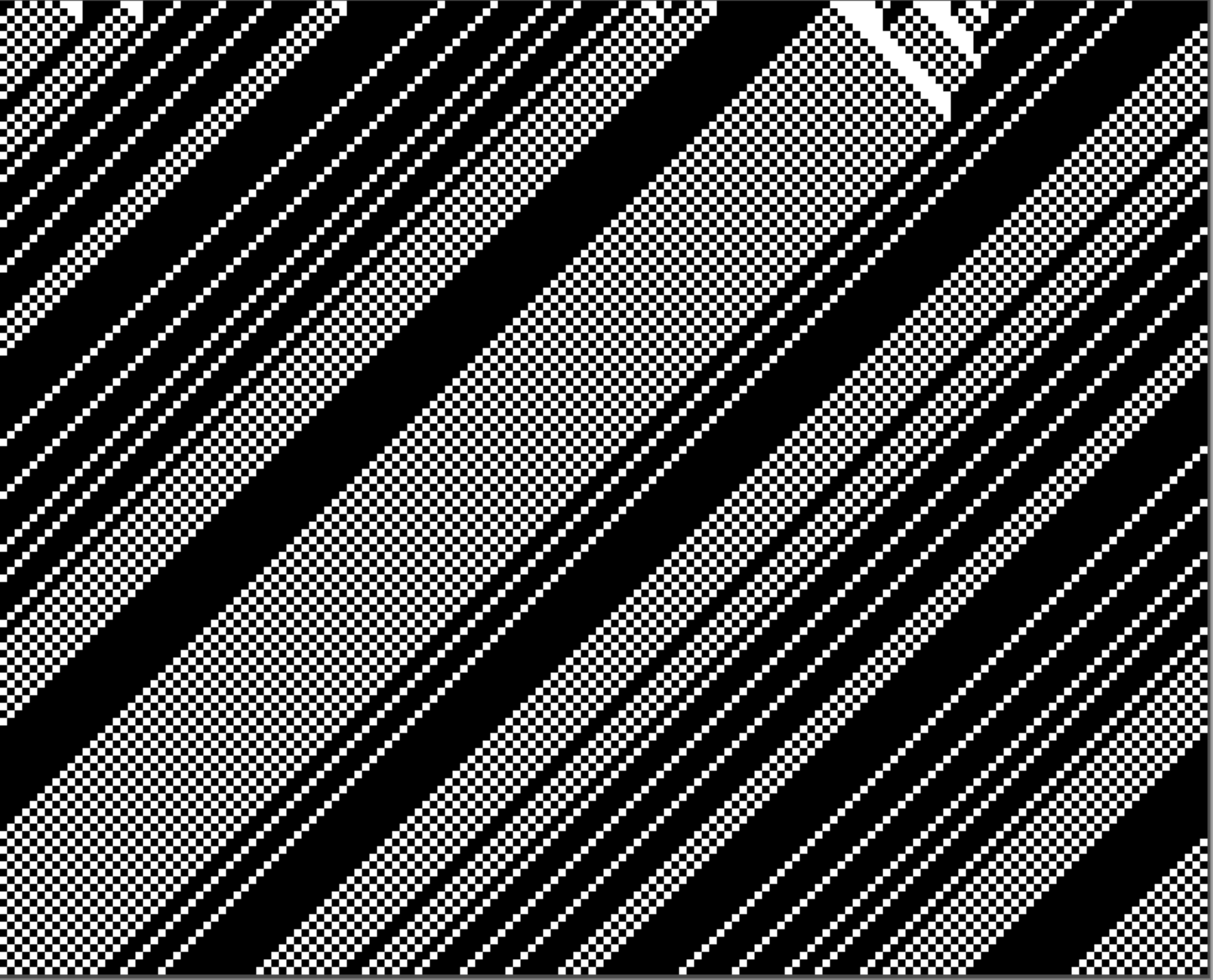}
	}

     \caption{Evolution of rule 184. Black cells (1) represent vehicles, white cells (0) represent spaces. Traffic flows to the right, time flows to the bottom: (A) In the \emph{free-flow} phase (density $\rho \leq 0.5$, $\rho=0.25$ shown) all vehicles flow at a velocity of one cell per tick (time step). (B) In the \emph{jammed} phase ($\rho > 0.5$, $\rho=0.75$ shown) jams move to the left, as vehicles can only advance when there is a free space ahead of them.}
     \label{fig:184}
\end{figure}

Elementary cellular automata (ECA)~\citep{Wolfram1986,WuenscheLesser1992,Wolfram:2002} are Boolean one dimensional CA, i.e.\ cells can take values 0 or 1 and are arranged in a one dimensional array, i.e.\ each cell has only two nearest neighbors. The state of  a cell at time $t$ depends on its state and the state of its nearest neighbors at time $t-1$. Thus, the state of each cell is determined by the states of three cells. There are $2^3=8$ possible combinations of values (0 or 1). The ECA ``rule" is a lookup table that determines the future state of cells depending on three cells. Thus, there are 256 possible rules, although many of them are somehow equivalent. In practice there are 88 equivalence classes~citep{WuenscheLesser1992}. Rule 184 (shown in Table~\ref{table:ECArules}) happens to model highway traffic flowing to the right. If there is space to the right, vehicles move there. Otherwise, they stay in their current cell. The temporal evolution of rule 184 is shown in Figure \ref{fig:184}.

\begin{figure}[htbp]
\begin{center}
\includegraphics[width=.6\textwidth]{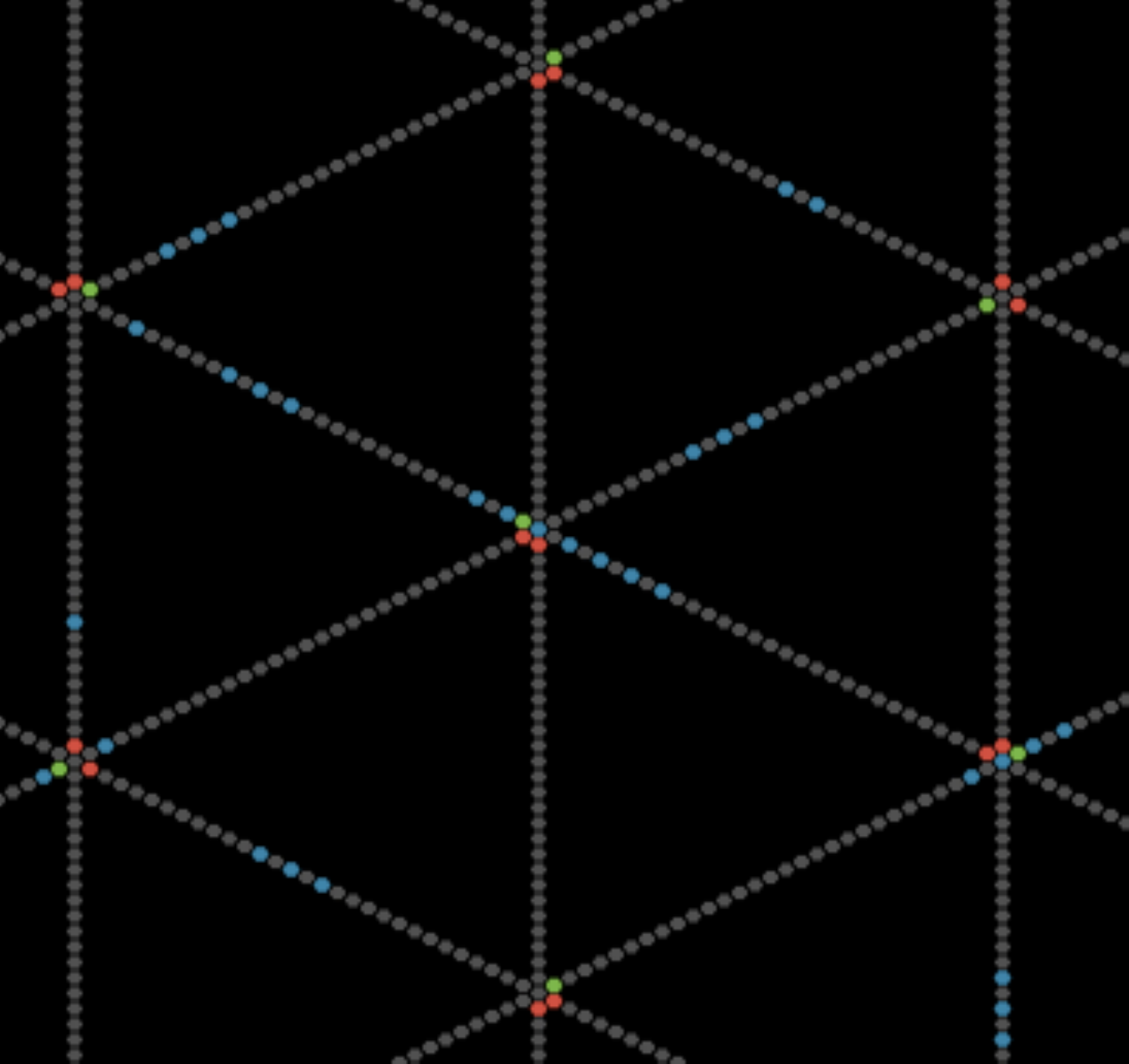}
\caption{Screenshot segment of hexagonal grid (density $\rho = 0.05$). \textcolor{sky}{Sky blue} cells represent vehicles, while \textcolor{darkgray}{dark gray} cells represent free spaces on streets. Black cells are not considered in the model.}
\label{fig:screenshot}
\end{center}
\end{figure}

We have previously proposed a model of city traffic based on elementary cellular automata~\citep{RosenbluethGershenson:2010} on a square grid.
In this section, we briefly review the model and extend it to an hexagonal grid. This allows the modeling of three-way intersections. 
 We consider single lane streets with periodic boundaries that can go in six different directions. Figure \ref{fig:screenshot} shows a screenshot segment of a simulation implementing our city traffic model on an hexagonal grid.
 
The behavior of vehicles on streets is already modeled with rule 184. Depending on how neighbors are chosen in the hexagonal grid, the direction of the streets can be set. To model intersections and traffic lights, we consider several coupled non-homogeneous ECA, where rules change around the intersection, depending on the state of the traffic light. The CA system is conservative~\citep{Moreira:2003}, i.e.\ the number of vehicles (1's) remains constant.


If a street has a green light, all its cells use rule 184. If there is
a red light, all cells on the street also use rule 184, with two exceptions: The
cell immediately before the intersection has to stop traffic from going
into the intersection. This is achieved with rule 252. The cell
immediately after the intersection
has to allow vehicles to leave, but not to allow vehicles in the intersection (flowing in a different direction) to enter the cell. This is achieved by rule 136. Table \ref{table:ECArules} lists the transition tables for the three rules used by the model.

\begin{figure}[htbp]
\begin{center}
\includegraphics[width=15cm]{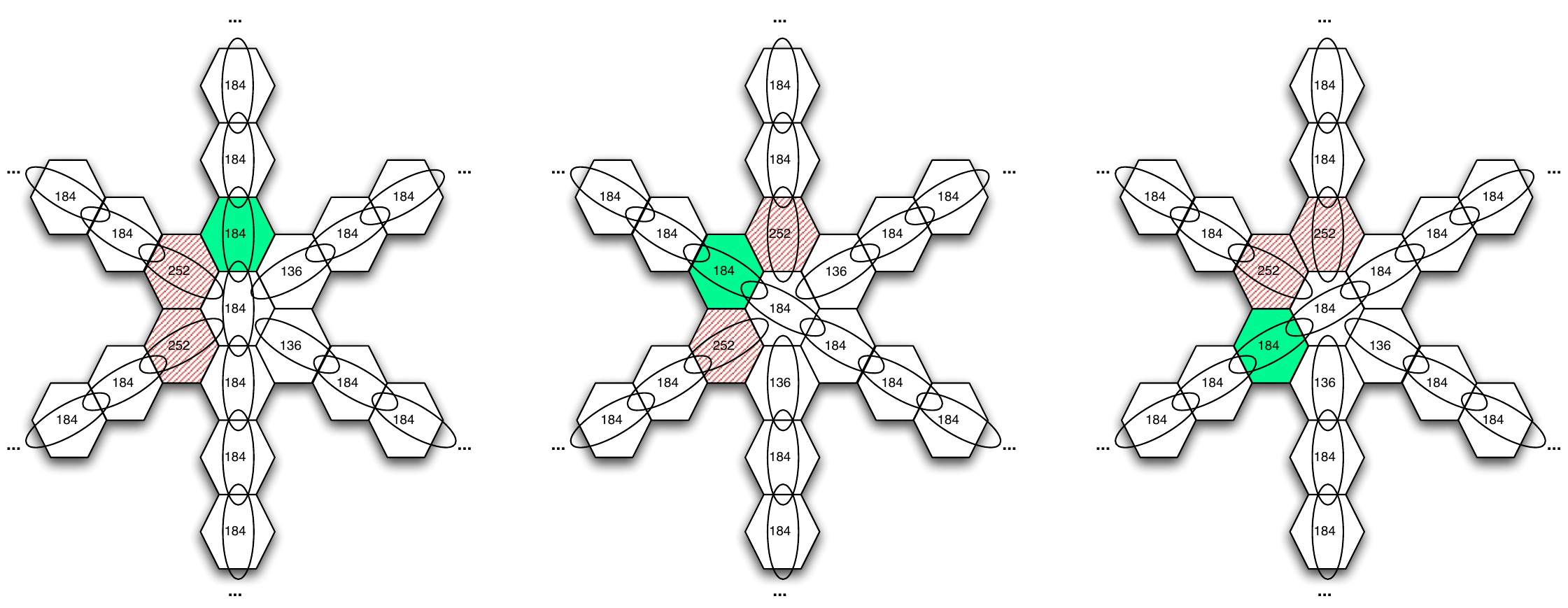}
\caption{Diagram for different rules (shown within cells) and
  neighborhoods (indicated by ovals) used around intersections,
  depending on the state of the traffic light. For green lights
  (indicated by \textcolor{green}{green} solid cells), rule 184 is used, and the
  intersection cell has as neighbors cells in the street with the
  green light. For red lights (indicated by  diagonally \textcolor{red}{red} striped
  cells), rule 252 is used for the cell immediately before the
  intersection and rule 136 for the cell immediately after the intersection. The rest of the cells use rule 184.}
\label{rulesDiagram-single}
\end{center}
\end{figure}

The intersection cell is a special case, as it has six 
neighbors. The rule never changes (184). What changes is the
neighborhood, i.e.\ it takes as nearest neighbors only the two cells in the street with a green light (also using rule 184). A diagram of the cells around an intersection is shown in Figure  \ref{rulesDiagram-single}.

\begin{table}[htdp]
\caption{ECA rules used in model}
\begin{center}
\begin{tabular}{|c|c|c|c|}
\hline
$t-1$	&$t_{184}$	&$t_{252}$	&$t_{136}$\\
\hline
000	&0	&0	&0	\\
\hline
001	&0	&0	&0	\\
\hline
010	&0	&1	&0	\\
\hline
011	&1	&1	&1	\\
\hline
100	&1	&1	&0	\\
\hline
101	&1	&1	&0	\\
\hline
110	&0	&1	&0	\\
\hline
111	&1	&1	&1	\\
\hline
\end{tabular}
\end{center}
\label{table:ECArules}
\end{table}%

If at time $t$ a traffic light is meant to switch, the model needs to
ensure that the intersection cell is empty. Otherwise, the vehicle in
the intersection would ``turn" into the crossing street. To avoid this
situation, the actual switching of rules and neighborhood is made only when the intersection is cleared.

\subsection{Measures}

The behavior of the model will depend strongly on the vehicle density
$\rho\in[0,1]$. The density can be easily calculated by dividing the number of cells with a 1 (i.e.\ total number of vehicles, $\sum{s_i}$) by the total number of cells ($\left|{S}\right|$):

\begin{equation}
\rho=\frac{\sum{s_i}}{\left|{S}\right|}
\label{eq:rho}
\end{equation}

The performance of the system can be measured with velocity $v\in[0,1]$, which is simply the number of cells that changed from 0 to 1 over the total number of vehicles:

\begin{equation}
v=\frac{\sum{{(s_i' > 0)}}}{\sum{s_i}}
\label{eq:v}
\end{equation}
where ${s_i'}$ is the derivative of state $s_i$.

The flux of the system represents how much of the space is used by moving vehicles. It can be obtained by multiplying the vehicle density by the velocity:

\begin{equation}
J=\rho v
\label{eq:J}
\end{equation}

The vehicular flow capacity of a triple intersection ($J_{max}=1/6$)
is 50\% lower than that of a double intersection ($J_{max}=1/4$), since intersection
time has to be shared with a third street. 

\subsection{Theoretical Optima}

When coordinating traffic lights, the best performance that can be theoretically achieved would be a system in
which each intersection has the best performance possible of an isolated intersection. A lower performance implies that there is \emph{interference} between traffic lights. Given the properties of our model of city traffic, equations \ref{eq:voptim} and \ref{eq:Joptim} describe the optimal curves for single intersections of velocity and flux, that depend on the maximum flux $J_{max}$ allowed by an intersection :

\begin{equation}
v_{optim} = \left \{ 
\begin{matrix}
1 & \mbox{if } \rho \leq J_{max} 
\\
\frac{1-\rho}{\rho} & \mbox{if } 1- J_{max} \leq \rho
\\
J_{max}/\rho & \mbox{if } J_{max} < \rho < 1- J_{max} 
\end{matrix}\right.
\label{eq:voptim}
\end{equation}

\begin{equation}
J_{optim} = \left \{ 
\begin{matrix}
\rho & \mbox{if } \rho \leq J_{max}   
\\
1-\rho & \mbox{if } 1- J_{max} \leq  \rho
\\
J_{max} & \mbox{if } J_{max} < \rho < 1- J_{max} 
\end{matrix}\right.
\label{eq:Joptim}
\end{equation}

If $ \rho \leq J_{max} $, then the intersection can support a maximum velocity of 1 cells/tick. Following equation \ref{eq:J}, the flux will then be equal to the density $\rho$, as all vehicles are moving. If $J_{max} < \rho < 1- J_{max}$, then the flux of the intersection will be restricted by the maximum capacity of the intersection, i.e. $J_{max}$. This implies that vehicles will be using the intersection at all times, and the average velocity will be $J_{max}/\rho$. If $1- J_{max} \leq \rho$, the density of the streets is so high that it restricts the flow of vehicles on streets, reducing the flux to $1-\rho$ and the velocity to $\frac{1-\rho}{\rho}$.

Notice that the flux optimum $J_{optim}$ is symmetric, since there is a symmetry in our city traffic model between vehicles (1's) moving in one direction and spaces (0's) moving in the opposite direction. This is also observed in the rule 184 model of highway traffic \citep{Kanai:2010}.

The ``interference" $\varPhi$ between traffic lights can thus be measured with the difference of the integrals of the optimal and experimental curves:

\begin{equation}
\varPhi_{v}=\int_{\rho_{min}}^{\rho_{max}}{v_{optim} - v}
\label{eq:Phiv}
\end{equation}
for velocity, and
\begin{equation}
\varPhi_{J}=\int_{\rho_{min}}^{\rho_{max}}{J_{optim} - J}
\label{eq:PhiJ}
\end{equation}
for flux.

Note that these equations are not normalized, i.e. different values of $J_{max}$ will yield different values for $\int{v_{optim}}$ and $\int{J_{optim}}$.

The Greek capital letter $\varPhi$ was chosen because the proposed measure of interference is related to the concept of ``friction" $\varphi$ \citep{GershensonDCSOS,Gershenson:2010a}, which measures the negative interaction between components of a system. This was proposed in the context of a general methodology for the design and control of self-organizing systems. Interference $\varPhi$ is a global average measuring different negative interactions between all intersections that affect the system performance.

\subsection{Scales}
\label{scales:subsect}
Even when the time and space are abstract and discrete, it can be assumed
that one cell represents five meters, roughly the space occupied by a vehicle. Thus, one kilometer of a street is represented by
200 cells. If each tick, i.e.\ time step, represents one third of
a second, then a velocity of one cell per tick is equivalent to 15
m/s, i.e.\ 54 km/h, equivalent to the speed limit in cities. A maximum
density of $\rho=1$ is equivalent to 200 vehicles per kilometer.

\section{Methods for coordinating traffic lights}
\label{sec:methods}

We briefly present in this section two methods for controlling traffic lights, that are more fully described in~\citet{GershensonRosenblueth:2010}.

The optimal coordination of traffic lights is an EXP-complete problem~\citep{PapadimitriouTsitsiklis1999,Lammer:2008}. This implies that it
is intractable. There have been several methods proposed to solve this
problem. We can distinguish two main approaches. One establishes fixed periods and phases that would maximize the flux for an expected traffic flow~\citep{FHA2005,Robertson:1969,N.H.Gartner:1975,SCATS1980,TorokKertesz1999,BrockfeldEtAl2001}. 
The other one tries to \emph{adapt}---manually or automatically---periods and phases
depending on current traffic flows~\citep{FHA2005,Henry:1983,Mauro:1990,Robertson:1991,FaietaHuberman1993,Gartner:2001,Diakaki:2003,FouladvandEtAl2004a,Mirchandani:2005,Bazzan2005,HelbingEtAl2005,Gershenson2005,CoolsEtAl2007,
GershensonRosenblueth:2010}.
A third conceivable approach would simply change the lights with fixed periods but random phases, i.e. no coordination between traffic lights.
We implemented three methods, one corresponding to each approach: a
\emph{green-wave} method that tries to maximize the flux by controlling
the phases, for an expected
traffic flow, a \emph{self-organizing} method that adapts to the
current traffic conditions, and a \emph{random} method.

\subsection{The \emph{green-wave} method}
\label{sec:gw}

The idea behind the \emph{green-wave} method~\citep{TorokKertesz1999} is the following: if the consecutive traffic lights switch with an offset (i.e.\ delay) equivalent to the expected vehicle travel time between intersections, vehicles should not have to stop. Thus, waves of green light move through the street at the same velocity as vehicles. This is the most commonly used method for coordinating traffic lights.

This method has advantages, e.g.\ when most of the traffic flows in
the direction of the green wave at low densities. On an hexagonal grid, only one direction can have vehicles without stopping ($v=1$). Vehicles on two other directions stop briefly each time around the cyclic boundaries. Vehicles on the three other directions have to stop considerably.
Moreover, if traffic is flowing at
velocities lower than expected, the green waves will go faster than vehicles and these will be delayed.

\subsection{The \emph{self-organizing} method}
\label{sec:so}

With the \emph{self-organizing} method, each intersection
independently follows the same set of rules, based only on local traffic information. There are only six rules (not related to ECA rules), with higher-numbered rules overriding lower-numbered ones. The full rule set is given in Table~\ref{table:rules}. This method is an improvement over previous work. The one reported in~\citep{Ball:2004} considered only rules 1 and 2, while~\citep{Gershenson2005,CoolsEtAl2007}, considered only rules 1--3. For a detailed description of the model, please refer to~\citet{GershensonRosenblueth:2010}. Here the rules are generalized naturally for several incoming streets per intersection. Details of the algorithm are presented in Appendix \ref{app:sola}.

\begin{table}[htdp]
\caption{\emph{Self-organizing} traffic light rules, independently followed by each intersection. Inset: Schematic of a triple intersection, indicating for the horizontal street distances $d$, $r$, and $e$ used for self-organizing lights.}
\begin{center}
\vspace{0.2cm}
\fbox{
\parbox{6 in}{

 \begin{center}
   \includegraphics[width=0.5\textwidth]{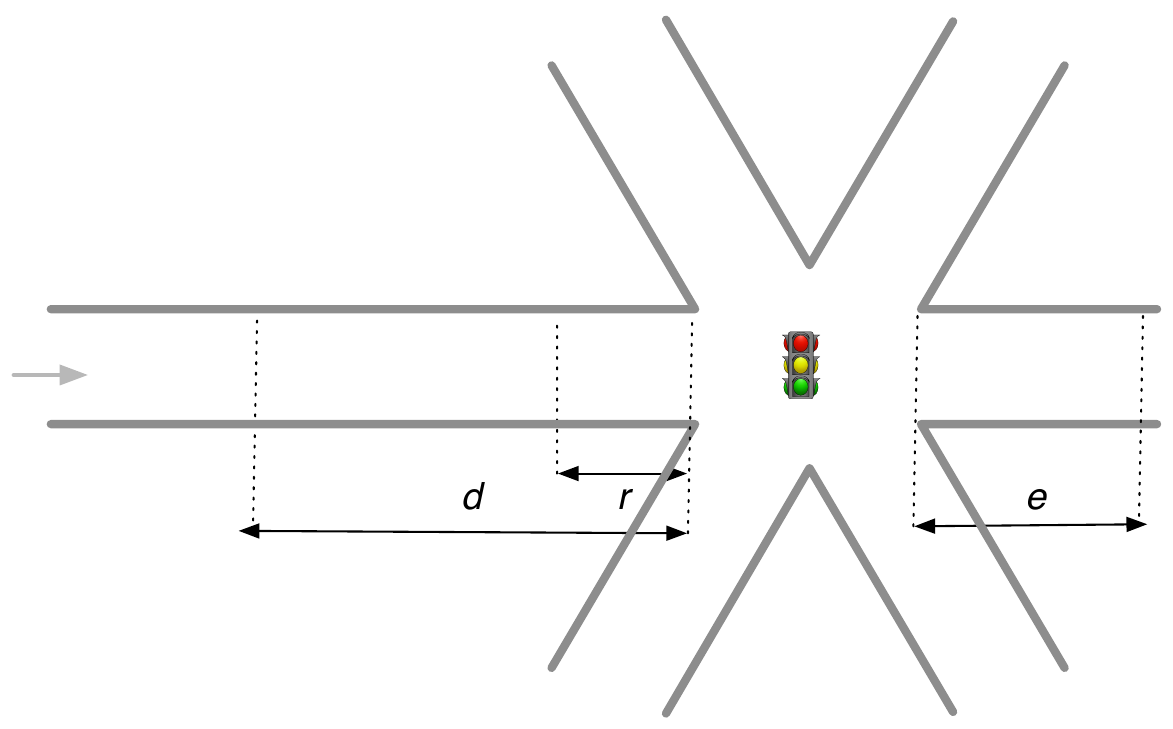}
 \end{center}

{\small
\begin{enumerate}
\item On every tick, add to counters $k_j$ the number of vehicles
approaching or waiting at red lights within distance $d$.
When the first counter exceeds a threshold $n$, switch the light.  (There are separate counters for each incoming direction $j$. Whenever the light switches, reset the counter for that direction to 0.)
\item Lights must remain green for a minimum time $u$.
\item If a few vehicles ($m$ or fewer, but more than zero) are left to cross a green light at a short distance $r$, do not switch the light.
\item If no vehicle is approaching a green light within a distance $d$, and at least one vehicle is approaching
a red light within a distance $d$, then switch that light to green.
\item If there is a vehicle stopped on the road a short distance $e$
  beyond a green traffic light, i.e.\ traffic is blocked upstream, then switch that light to red. Switch to green the direction with a highest value in counter $k_j$ and no blockage upstream.
\item If there are vehicles stopped on all directions at a short distance $e$
  beyond the intersection, then switch all lights to red. Once one of the directions is free, restore the green light in that direction.
\end{enumerate}

}

}

}

\end{center}
\label{table:rules}
\end{table}%

The main idea is the following: each intersection counts how many vehicles are behind red lights (approaching or waiting). Each time interval, the vehicular count is added to a counter which represents the integral of vehicles over time approaching the intersection on each direction with a red light. When one of these counters reaches a certain threshold, the red light switches to green (rule 1 in Table~\ref{table:rules}). If there are few vehicles approaching, the counter will take longer to reach the threshold. This increases the probability that more vehicles will aggregate behind those already waiting, promoting the formation of ``platoons". The more vehicles there are, the faster they will get a green light. Like this, platoons of a certain size might not have to stop at intersections. There are other simple rules to ensure a traffic smooth flow. The method adapts to the current traffic density and responds to it efficiently: For low densities, almost no vehicle has to stop. For medium densities, intersections are used at their maximum capacity, i.e.\ there are always vehicles crossing intersections, there is no wasted time. For high densities, most vehicles are stopped, but gridlock is avoided with simple rules that coordinate the flow of ``free spaces" in the opposite direction of traffic, allowing vehicles to advance (rules 5 and 6).

Intersections with only two incoming streets will have one green and one red light, or both lights red. The algorithm was extended naturally for more than two incoming streets. Only one street (or none) will have a green light. In this way, every direction $j$ with a red light will keep a counter $k_j$ quantifying incoming vehicles. The first one to reach the threshold will request the green light and be reset. Other directions will not reset their counter $k_j$, so they will request a green light soon if there is the demand for it. For other rules, the street with highest demand (measured with $k_j$) will usually get the preference, unless traffic is blocked upstream. Notice that if there are e.g. three incoming directions A, B, and C, there is no prefixed sequence of which direction will get a green light. This will depend entirely on the actual traffic conditions. An example switching pattern could be A, B, A, C, B, A, C, B, C, B, A, B\ldots The periods are also variable (with the constraint of a minimum green time), and depend on the traffic demand. Each switching usually has a different green period.

\section{Three streets: one or three intersections?}
\label{sec:single}

\begin{figure*}
     \centering
     \subfigure[]{
          \label{fig:scenariosA}
          \includegraphics[width=.35\textwidth]{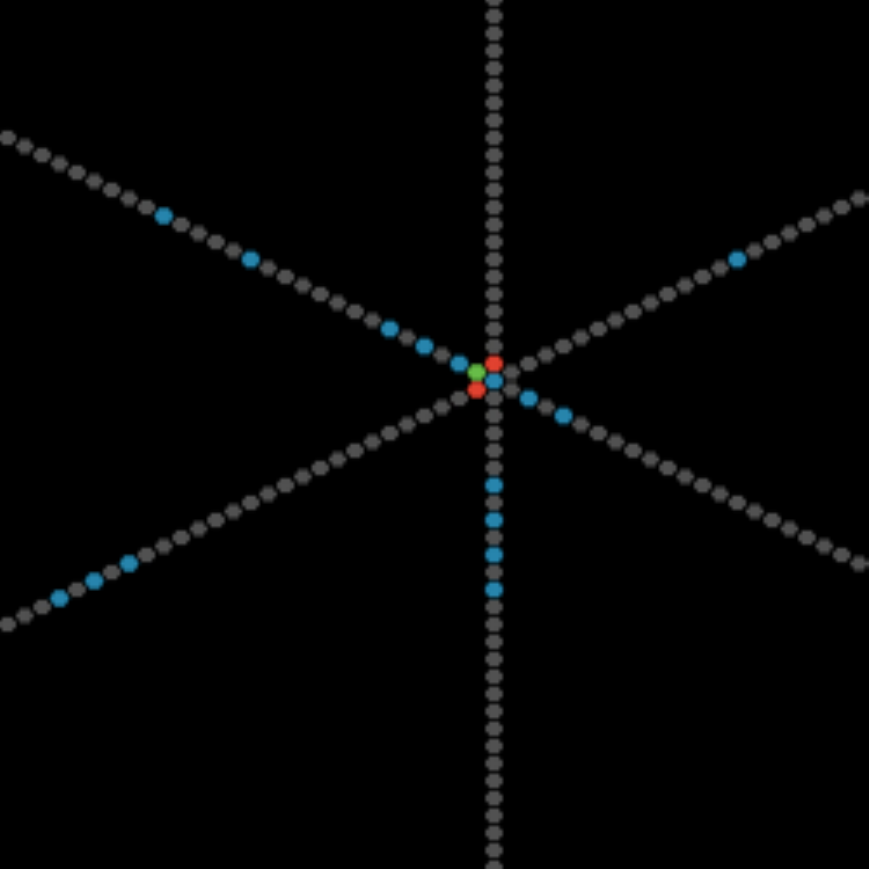}}
     \subfigure[]{
          \label{fig:scenariosB}
          \includegraphics[width=.35\textwidth]{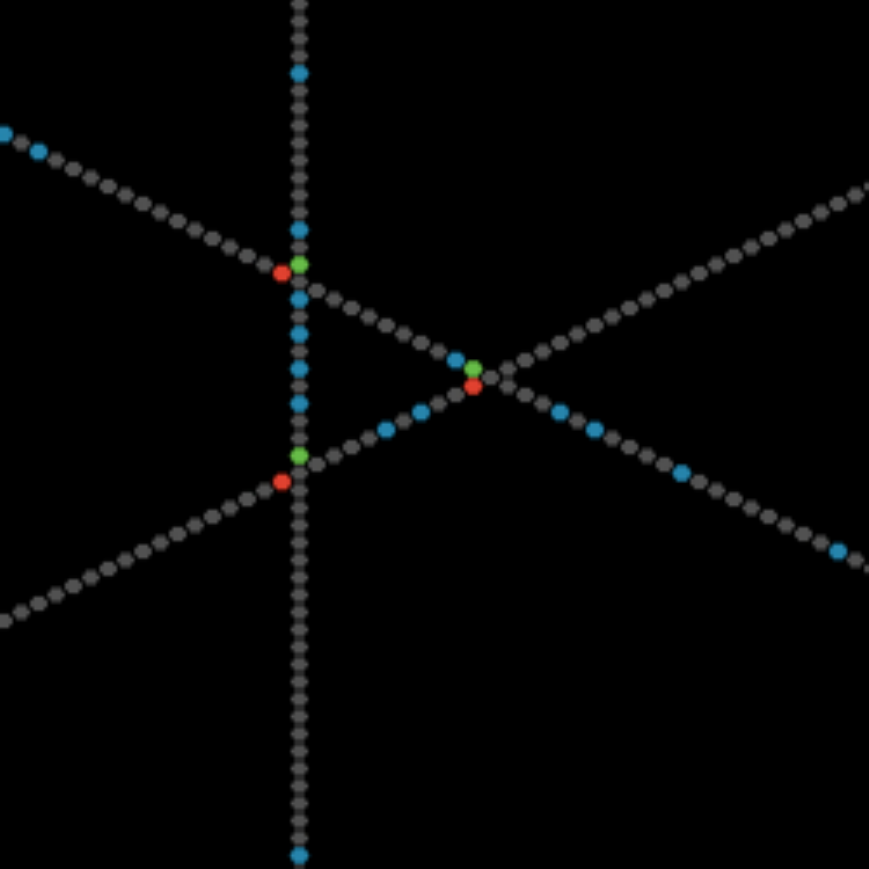}}

     \caption{Studied scenarios involving three streets: (A) one triple intersection and (B) three double intersections. For both scenarios, the length of cyclic streets is 180 cells. \textcolor{sky}{Sky blue} cells represent vehicles, while \textcolor{darkgray}{dark gray} cells represent free spaces on streets. }
     \label{fig:scenarios}
\end{figure*}

In this section, we compare a scenario with a single triple intersection with another scenario with three double intersections.
We developed
a computer simulation in NetLogo~\citep{Wilensky1999}. The reader is invited to access the simulation via web browser at the URL \url{http://turing.iimas.unam.mx/~cgg/NetLogo/4.1/trafficHexCA.html} (for short, \url{http://tinyurl.com/tHexCA}).
The environment consists of three cyclic streets, i.e.\ with periodic boundaries. Each street has a length of 180 cells. We explore two scenarios: one where the three streets meet at a single cell, i.e.\ one triple intersection, and another once consisting of three double intersections with a distance of 11 cells between intersections. A section of the simulation showing the intersection arrangements of both scenarios can be seen in Figure \ref{fig:scenarios}.

\begin{figure}
     \centering
     \subfigure{
          \label{fig:results_singleA}
          \includegraphics[width=.45\textwidth]{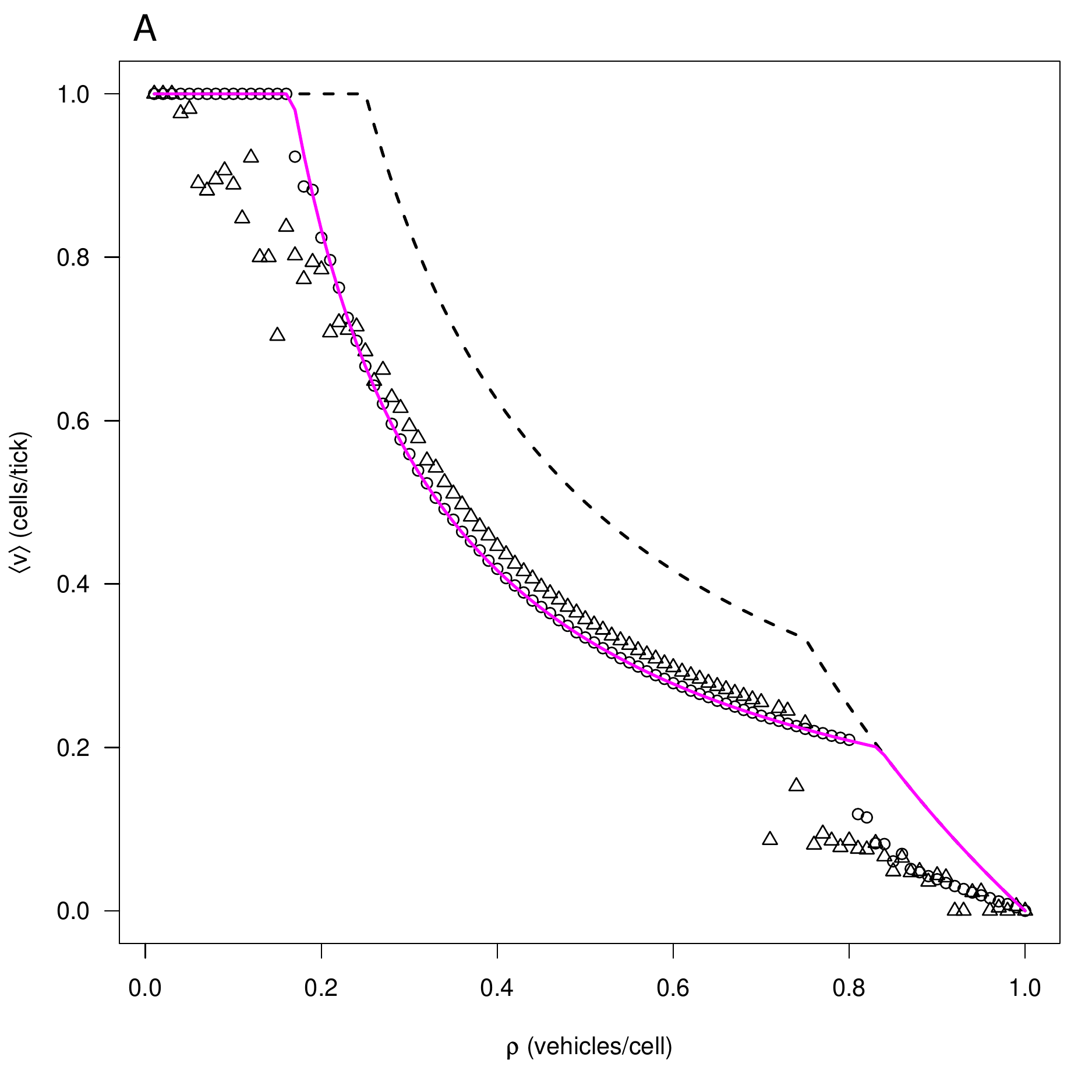}}
     \subfigure{
          \label{fig:results_singleB}
          \includegraphics[width=.45\textwidth]{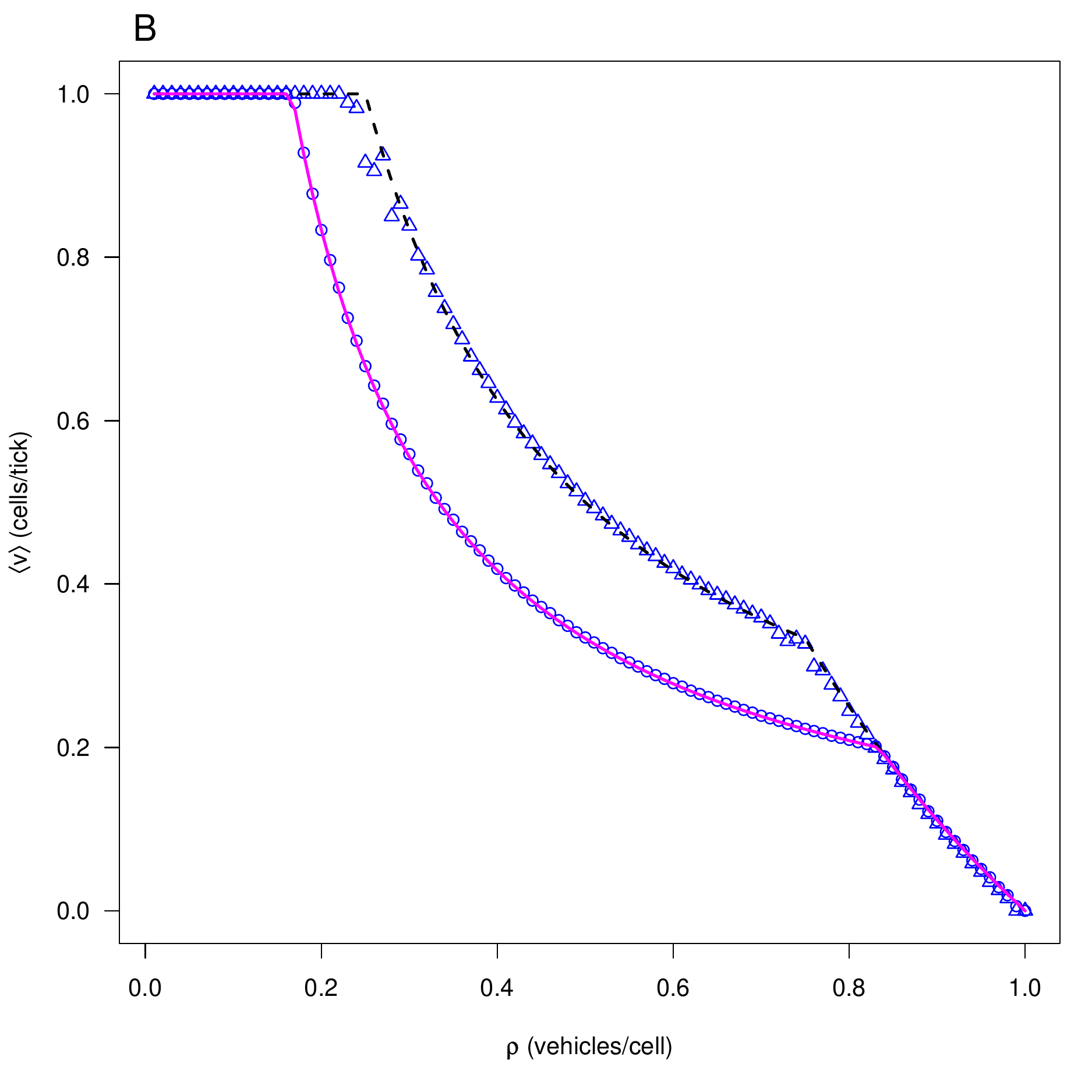}}
\\
     \subfigure{
          \label{fig:results_singleC}
          \includegraphics[width=.45\textwidth]{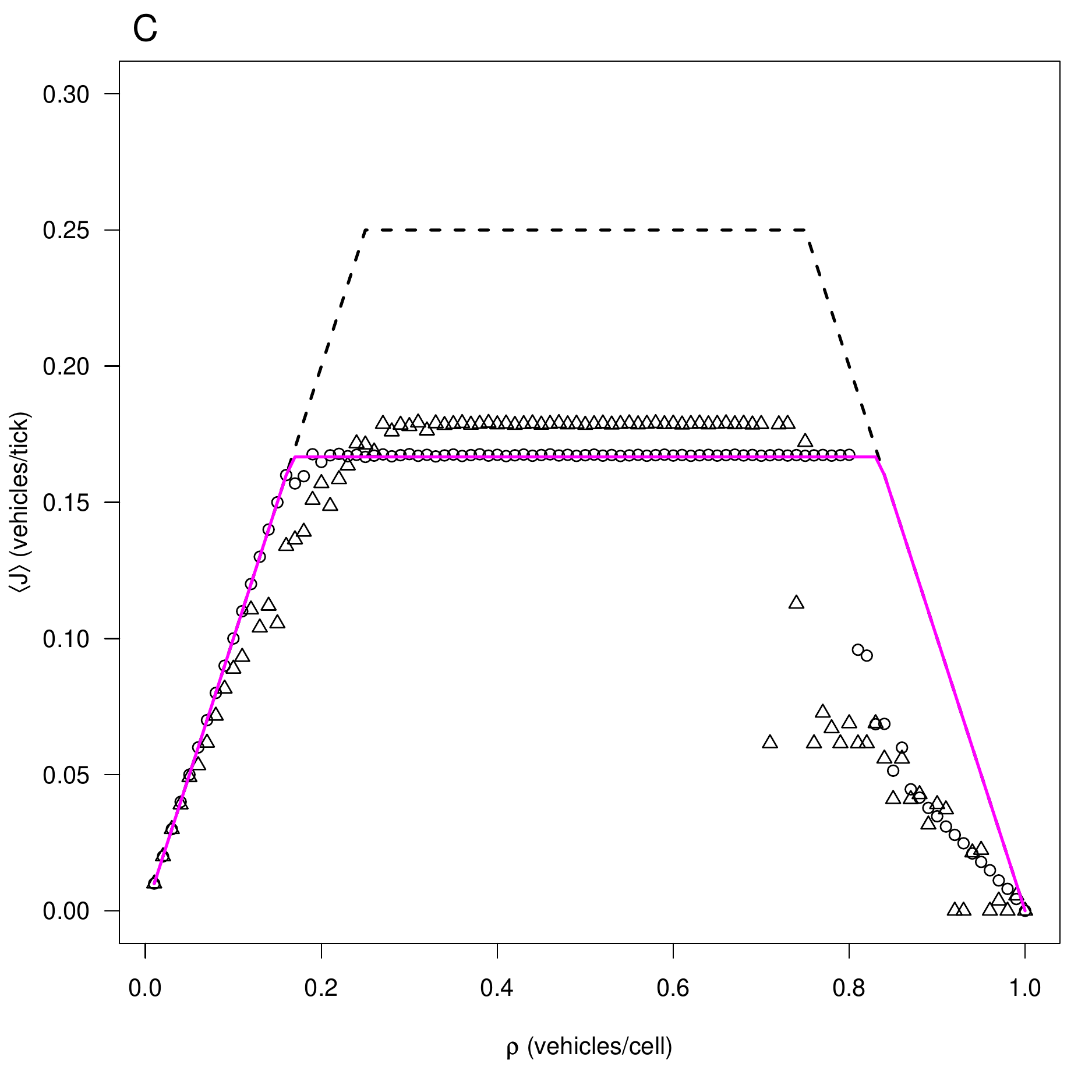}}
     \subfigure{
          \label{fig:results_singleD}
          \includegraphics[width=.45\textwidth]{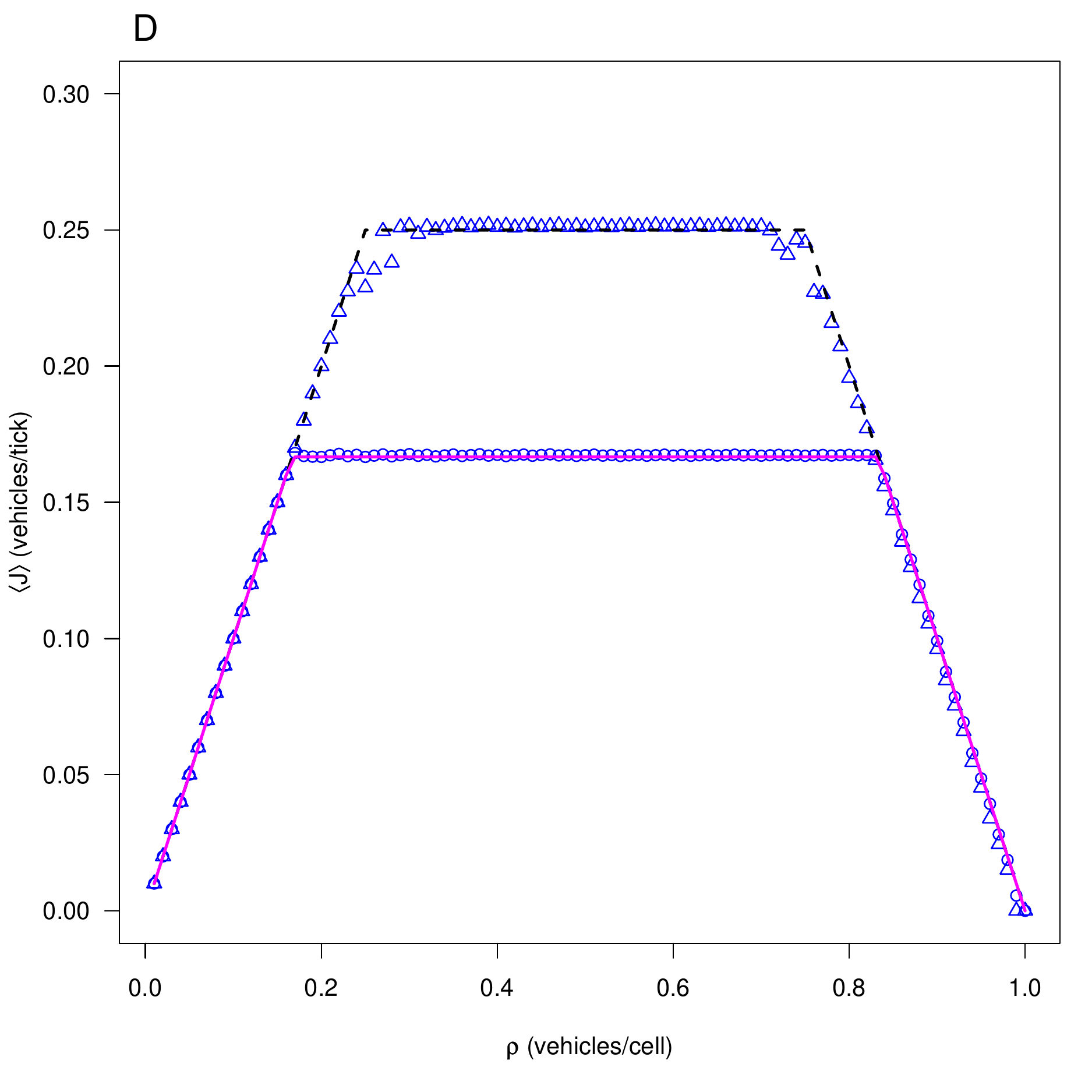}}

     \caption{Simulation results for a single intersection: (A,B) average velocity $\langle v\rangle$  and (C,D)  average flux $\langle J\rangle$ for different densities $\rho$. Black points indicate fixed traffic lights (A,C), while blue points indicate self-organizing traffic lights (B,D), in scenarios with one triple intersection ({\Large $\circ$}) and  three double intersections ($\bigtriangleup$). Optimality curves are shown with a \textcolor{magenta}{magenta} continuous line (for single triple intersection scenario) and a with black dashed line (for three double intersections scenario).}
     \label{fig:results_single}
\end{figure}

For the experiments, each run consisted of the following: Half an hour
of virtual time (as opposed to real time),
corresponding to 5,400 ticks (see subsection~\ref{scales:subsect}) was
simulated for random initial conditions. Since the 
vehicles are placed randomly, one street may have a slightly higher
density than another. After this initial half an hour of virtual time, 
the system is considered to have stabilized, i.e.\ gone through a
transient, so 
another half an hour is simulated, of which the average velocity of all vehicles is measured at
every tick (see equation \ref{eq:v}). At the end of the simulation, the velocities of the second
half an hour are averaged to obtain the average velocity $\langle
v\rangle$ and average flux $\langle J\rangle$ for a given density $\rho$ (see equation \ref{eq:J}). The results are shown
in Figure \ref{fig:results_single}. Since successive simulations show
low standard deviations---apart from phase transitions---only single
runs per density are shown for clarity.

For both scenarios, we compared a \emph{fixed-period} method with the
\emph{self-organizing} method. The fixed method used a period of
$T=180$ ticks. For the triple intersection scenario, this period
$T=180$ implies that each direction will have a green phase of 60
ticks. Moreover, since the length of streets is equal to the period, a
vehicle can in principle \emph{free flow} in any direction,
i.e.\ $v=1$. This is the case for densities
$\rho\leq 1/6$. Afterwards, there is no possibility to let all
vehicles flow freely, so there is a phase transition into
\emph{intermittent} flow, where some vehicles have to stop. This phase
is characterized by a maximum flux $J=1/6$, given by the triple
intersection. After a density of $\rho \approx 0.8$, the waiting
queues grow long enough to block the intersection while another street
should have a green light.  This is noticeable in the abrupt decrease
in velocity and flux. We call this phase ``\emph{interfered}", since
the queue of one street interferes with and blocks another street. 

For the scenario with the three double intersections, the period
$T=180$ implies that each direction will have a green phase of 90
ticks. This increases the capacity of the intersections by
50\%. 
However, the coordination of the traffic lights now has an effect on
the flux.
In these simulations, all three traffic lights changed at
the same time, i.e.\ there was no green wave. This restricts the \emph{free
flow} phase for very small densities ($\rho \lesssim 0.05$,
depending on initial conditions). 
The \emph{intermittent phase}
does not reach a maximum possible flux of $J=0.5$, although the flux
is slightly higher than for the triple intersection scenario. Still,
the transition into the \emph{interferred} phase occurs at a lower
density. 


For the \emph{self-organizing} method (subfigures
\ref{fig:results_singleB} and \ref{fig:results_singleD}), the scenario
with a single triple intersection performs for most densities as the
\emph{fixed-period} method, since there is no coordination of traffic
lights involved, and the maximum flux is determined by the
intersection, not so much by the controller method. There is a
\emph{free-flow} phase for $\rho\leq 1/6$, and then an
\emph{intermittent} phase until $\rho\leq5/6$. However, for high
densities ($\rho\geq5/6$), the intersection is not blocked due to
rules 5 and 6. Therefore, there is no abrupt decrease in flux. When a
street is blocked ahead, another one without blockage is given a green
light. Thus, for densities $\rho\geq5/6$ there is a
``\emph{quasi-gridlock}" phase, where most vehicles are stopped, but
flow is not stopped. In fact, ``free-spaces" move in the direction
opposite of the vehicles with a velocity close to $-1$ cells/tick. 

For the scenario with three double intersections, it can be seen that
the \emph{self-organizing} method manages to coordinate the traffic
flows, and the performance is very close to optimal. There is
\emph{free-flow} until $\rho\approx1/4$ and then \emph{intermittent}
flow until $\rho\approx3/4$. For higher densities ($\rho
\gtrapprox 3/4$), the \emph{interferred} phase is also avoided,
giving place to \emph{quasi-gridlock}. 

It is interesting to note that there is a symmetry in the flux diagram
for both scenarios. The flux diagram of the rule 186 model of highway
traffic is also symmetric~\citep{Kanai:2010}. This is given by the
fact that there is a certain duality between vehicles (1) and spaces
(0) in these models.

For the \emph{self-organizing} method, the flux is the same for both
scenarios at $J< 1/6$. 
This further suggests that the restrictions on traffic flow depend on the street topology and not on the \emph{self-organizing} method, which adapts to both scenarios. 

In the next section, we study more complicated scenarios, where a large
number of intersections has to be coordinated. 

\section{18 streets: 36 or 108 intersections?}
\label{sec:city}

\begin{figure}[htbp]
\begin{center}
\includegraphics[width=.85\textwidth]{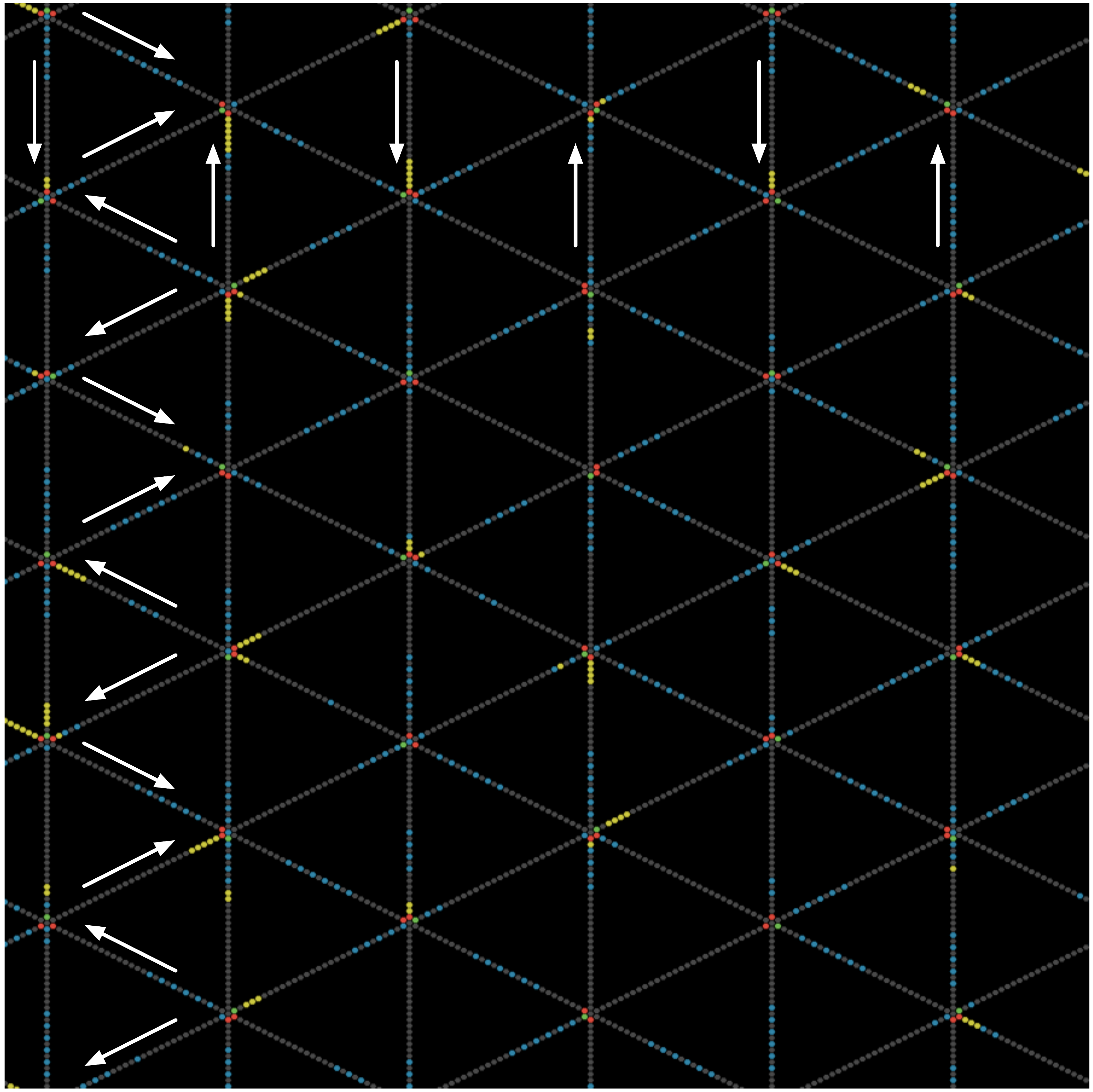}
\caption{Screenshot of scenario with 18 streets and 36 triple intersections, $\rho=1/5$. Moving vehicles are colored with \textcolor{sky}{sky blue} and stopped vehicles are colored in \textcolor{yell}{yellow}.}
\label{fig:36triple}
\end{center}
\end{figure}

\begin{figure}[htbp]
\begin{center}
\includegraphics[width=.85\textwidth]{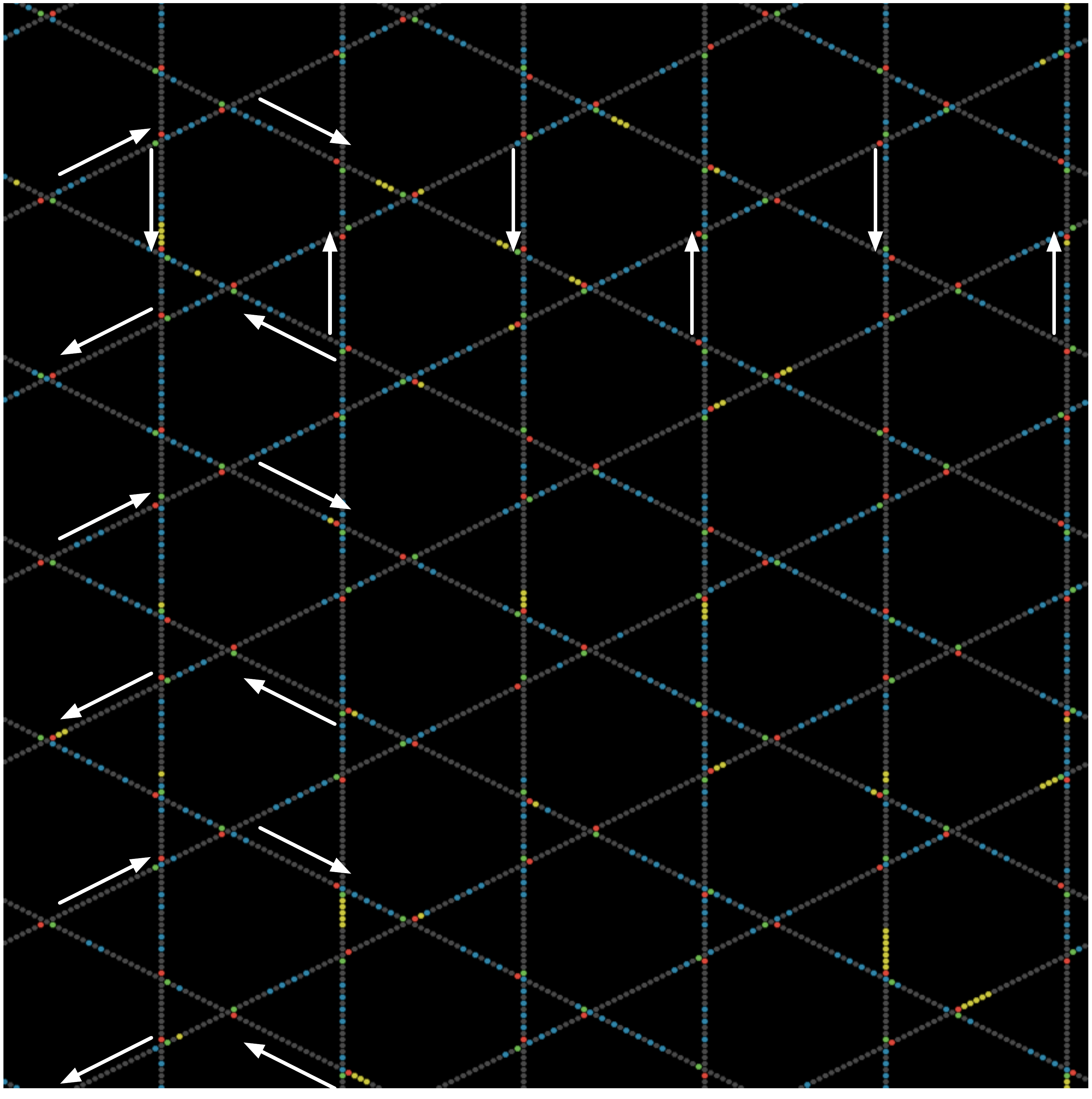}
\caption{Screenshot of scenario with 18 streets and 108 double intersections, $\rho=1/5$. Moving vehicles are colored with \textcolor{sky}{sky blue} and stopped vehicles are colored in \textcolor{yell}{yellow}.}
\label{fig:108double}
\end{center}
\end{figure}

To study the coordination of several traffic lights, we simulated 18
cyclic streets of 180 cells each, three of which have traffic flowing in each
of six directions. In one scenario, shown in Figure \ref{fig:36triple}, the streets were setup to cross at
36 triple intersections, having 30 cells between intersections,
i.e.\ one block is 30 cells. In a second scenario, shown in Figure \ref{fig:108double}, the setup was such
that streets crossed at 108 double intersections, with heterogeneous
block distances of 11 or 19 cells. 

\begin{figure}
     \centering
     \subfigure{
          \label{fig:results_cityA}
          \includegraphics[width=.6\textwidth]{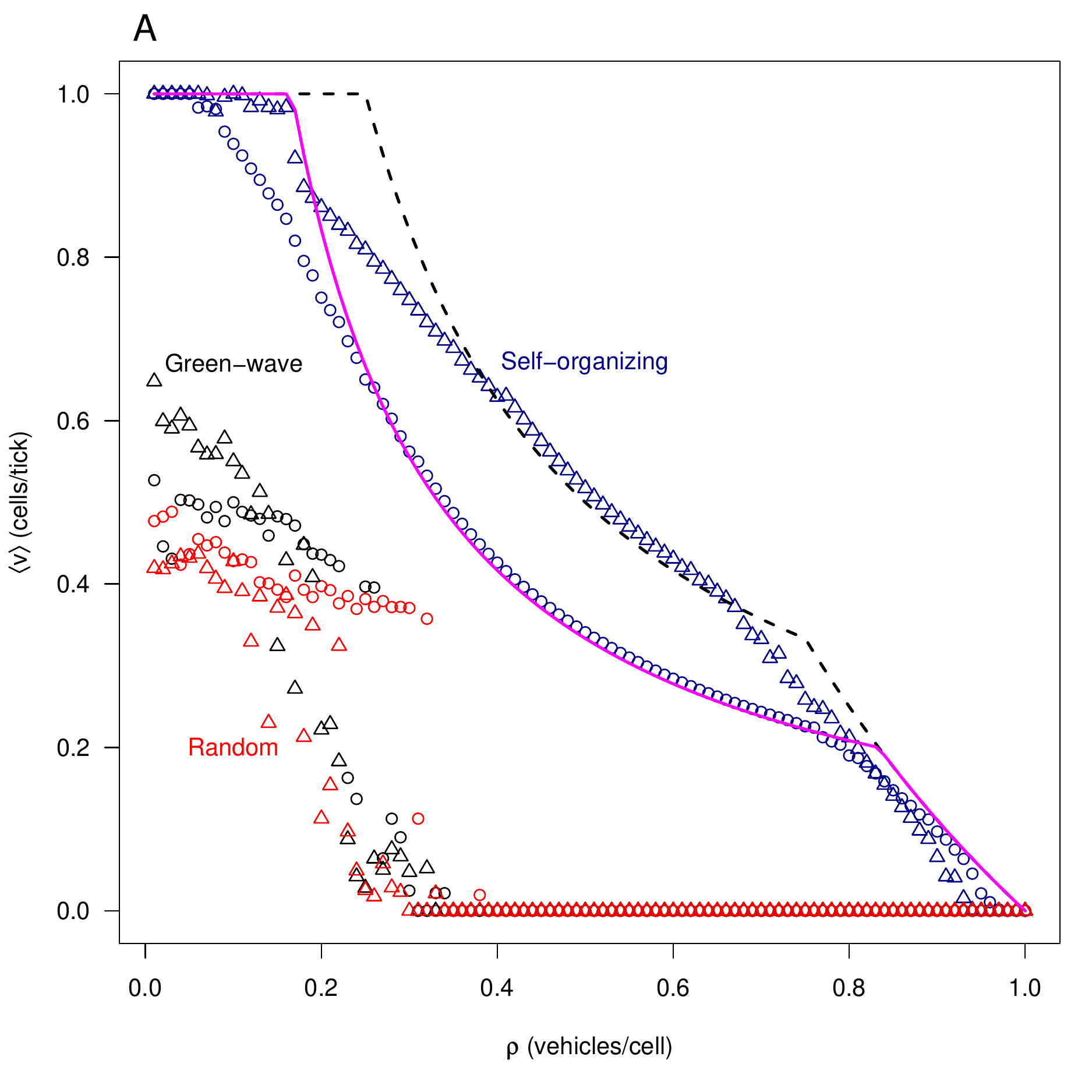}}
     \\     
     \subfigure{
          \label{fig:results_cityB}
          \includegraphics[width=.6\textwidth]{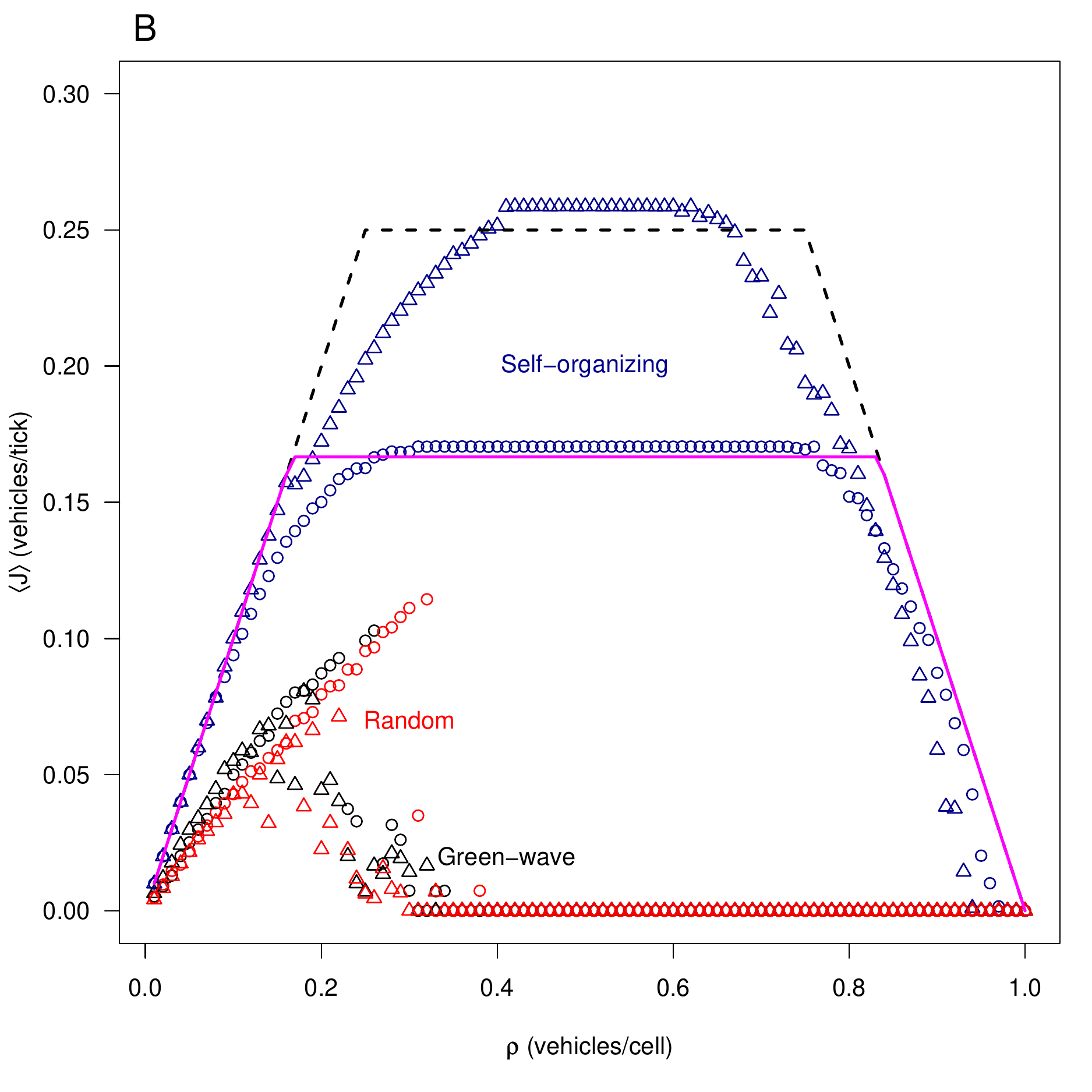}}

     \caption{Simulation results for 18 street scenarios: (A) average velocity $\langle v\rangle$ and (B) average flux $\langle J\rangle$ for different densities $\rho$, \emph{green-wave} (black), \emph{self-organizing} (\textcolor{blue}{blue}), and \emph{random} (\textcolor{red}{red}) methods. Two scenarios are considered: 36 triple intersections ({\Large $\circ$}) and 108 double intersections ($\bigtriangleup$). Optimality curves are shown with a \textcolor{magenta}{magenta} continuous line (for single triple intersection scenario) and a with black dashed line (for three double intersections scenario).}
     \label{fig:results_city}
\end{figure}

The same simulation developed in NetLogo and experimental setup as the one described in the previous section were used. The simulation is available at the URL \url{http://tinyurl.com/tHexCA}. Results are shown in Figure \ref{fig:results_city}. As a worst case scenario, a \emph{random} method is included in these simulations, where each traffic light has a fixed time period with randomly assigned offsets, i.e. no correlation between traffic lights.

It can be seen that the \emph{green-wave} method offers a poor performance. Having six directions, only one can be fully coordinated with the green wave. Two other directions are partially coordinated, but three are anti-coordinated with the green-wave. Even when theoretically the scenario with double intersections should be more efficient---given the fact that intersections have $50\%$ more capacity---the performance difference with the triple intersection scenario is minimal for low densities. Moreover, for densities $\rho \approx 0.2$, the performance in the triple intersection scenario is more efficient. Still, the \emph{green-wave} method collapses into gridlock for medium densities ($\rho \gtrapprox 0.3$). This is because traffic flowing against the green wave accumulate in long queues that block intersections upstream, eventually leading to partial or complete gridlocks. This performance is comparable with that of a \emph{random} method, where no traffic light is correlated.

For the \emph{self-organizing} method, at low densities ($\rho \leq 0.05$) free flow is reached in both scenarios, i.e.\ vehicles are coordinated in six directions and $v=1$, i.e.\ no vehicle has to stop. This is of course an artifact of the cyclic boundaries, but it illustrates the coordination capacity of the \emph{self-organizing} method. As the density increases, the average velocity decreases gradually (Figure \ref{fig:results_cityA}). Gridlock is reached only for very high densities, where initial conditions already block intersections.

There is a noticeable difference in the performance of the \emph{self-organizing} method in both scenarios. In both of them a maximum flux is reached according to the intersection type: $J=1/6$ for triple intersections and $J=1/4$ for double intersections (Figure \ref{fig:results_cityB})\footnote{The supraoptimal results ($J>J_{max}$) are an artifact of the city traffic model \citep{RosenbluethGershenson:2010}. Rule 184 assumes a $J_{max}=0.5$ for single streets since a free space is required between occupied cells for vehicles to flow. However, in intersections, this requirement can be relaxed once a light is switched, i.e. if a light changes, a vehicle can occupy an intersection the tick after a vehicle going on another street crossed the intersection. This slightly increases the analytical maximum intersection capacity, depending on the traffic light switching frequency.}. A good performance is achieved, restricted by the street topology, as opposed to the \emph{green-wave} method, where inefficient flow is produced independently of the street topology, i.e.\ it cannot exploit the increased capacity of double intersections because it cannot coordinate so many directions and intersections. In contrast, the \emph{self-organizing} method is able to coordinate many directions and has shown to be highly scalable. 

The triple intersection scenario is less deviated from the optimal curves, indicating that it is computationally easier to coordinate less intersections. Still, the double intersection scenario achieves a better performance, in spite of deviating more from the optimality curves, since it is more complicated to coordinate more intersections.

A clearer understanding of the performance of the \emph{self-organizing} method will be obtained by analyzing the dynamical phases that are found as the density changes.

\subsection{Dynamical Phases}

The \emph{green-wave} method has only two dynamical phases in both scenarios: \emph{intermittent} (some vehicles move, some are stopped) and \emph{gridlock} (all vehicles are stopped, $v=0$). 

\begin{figure}
     \centering
     \subfigure{
          \label{fig:phasesA}
          \includegraphics[width=.6\textwidth]{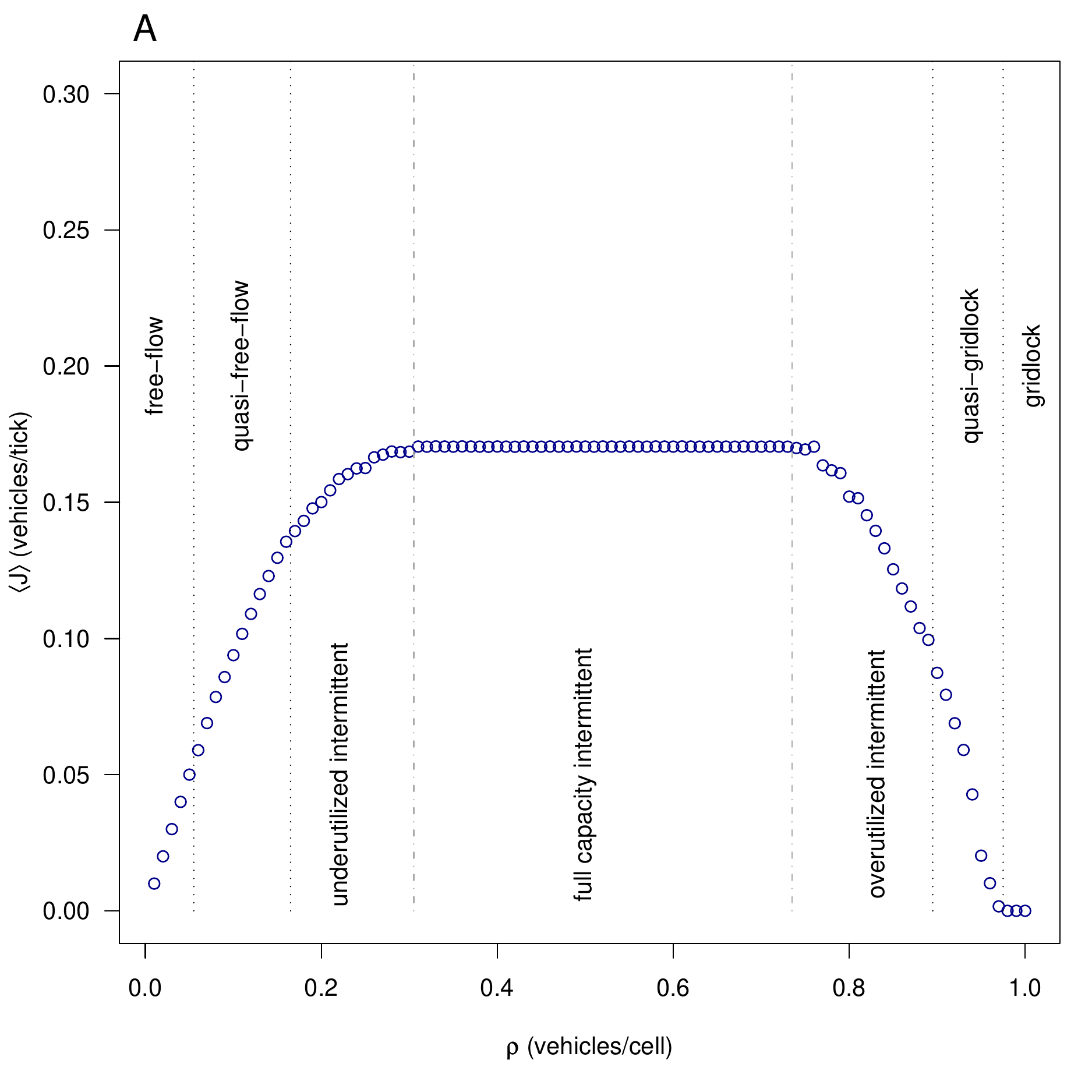}}
     \\     
     \subfigure{
          \label{fig:phasesB}
          \includegraphics[width=.6\textwidth]{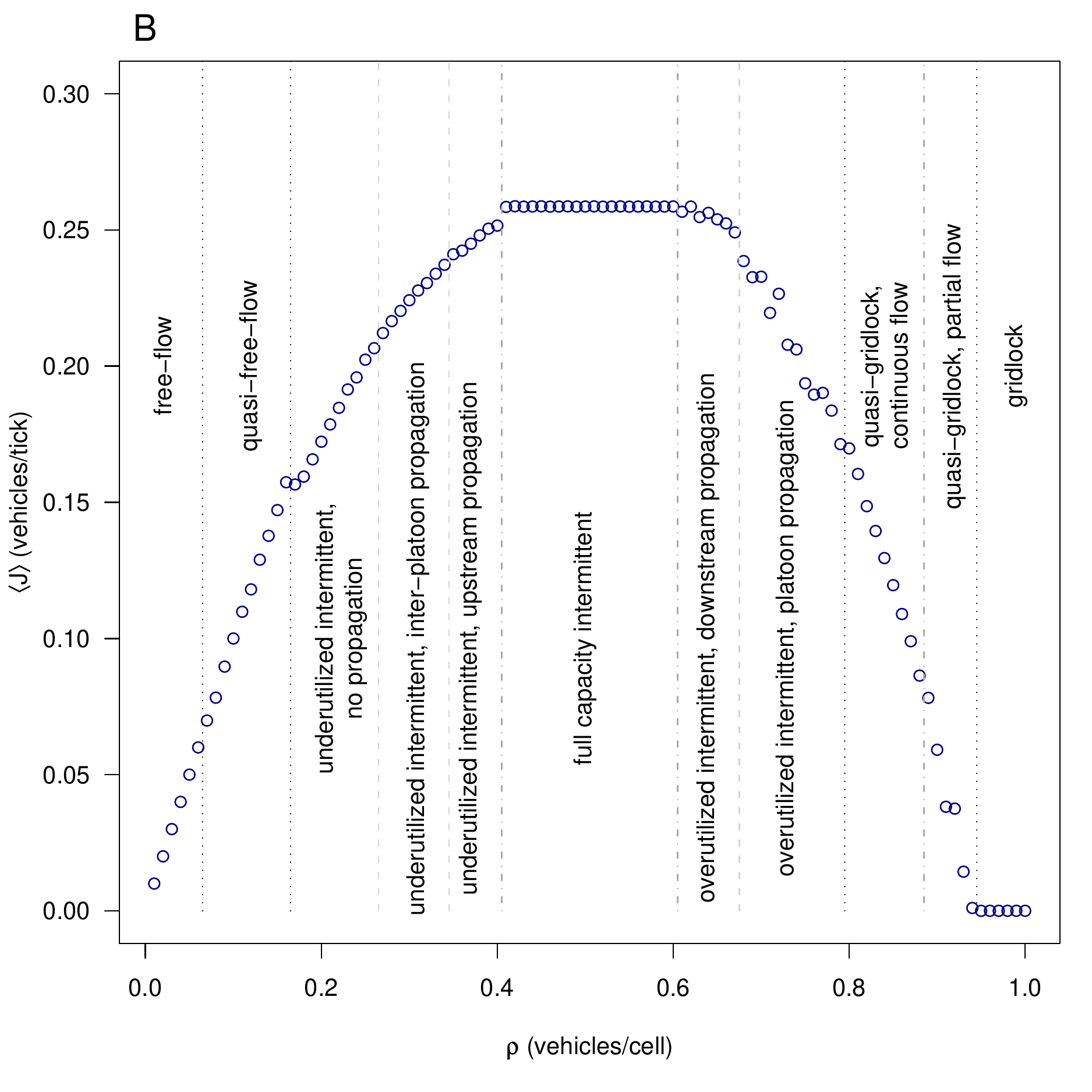}}

     \caption{Phase transitions shown in flux diagrams: (A) 36 triple intersections scenario and (B) 108 double intersections scenario.}
     \label{fig:phases}
\end{figure}

Phase transitions for the \emph{self-organizing} method are indicated in the flux diagrams for both scenarios in Figure \ref{fig:phases}.

For the triple intersection scenario (Figure \ref{fig:phasesA}), the same seven phases as with a square grid were observed~\citep{GershensonRosenblueth:2010}: \emph{free-flow}, \emph{quasi-free-flow}, \emph{underutilized intermittent}, \emph{full capacity intermittent}, \emph{overutilized intermittent}, \emph{quasi-gridlock}, and \emph{gridlock} (to be described below). Still, the phase transitions occur at different densities.

For the double intersection scenario (Figure \ref{fig:phasesB}), apart from the phases already detected, further subphases appeared, with a total of ten phase transitions.

\begin{description}
\item[Free-flow] ($\rho \leq 0.06$). In this phase, all vehicles have a maximum velocity $v=1$. This implies a perfect coordination between all six different flow directions.

\item[Quasi-free-flow] ($0.07 \geq \rho \geq 0.16$). Most vehicles are flowing, but some have to stop momentarily, when platoons coincide at a given time at an intersection. 

\item[Intermittent] ($0.17 \geq \rho \geq 0.79$). Some vehicles stop while others flow. It can be subdivided into three subphases: underutilized, full capacity, and overutilized.

\begin{description}
\item[Underutilized] ($0.17 \geq \rho \geq 0.4$). This phase is characterized by a partial utilization of the intersection's full capacity, i.e.\ $J<J_{max}$ ($J_{max}=1/6$ for triple intersections, $J_{max}=1/4$ for double intersections). $J_{max}$ implies that at every intersection vehicles are crossing all the time. The underutilized intermittent phase has a flux $J<J_{max}$ because there are not enough vehicles to reach $J_{max}$, i.e.\ intersections are not utilized at their maximum capacity. Still, flux $J$ is higher than the maximum flux $J_{max}$ for triple intersections. This subphase can be further subdivided in three subphases for the double intersection scenario: no propagation, inter-platoon propagation, and upstream propagation.
\begin{description}
\item[No propagation] ($0.17 \geq \rho \geq 0.26$). Queues form, but these do not propagate, i.e.\ after the simulation has stabilized, queues do not propagate between platoons: when a platoon has to wait behind a red light, it will quickly get a green light and its last vehicle will be moving before another vehicle reaches it and stops.
\item[Inter-platoon propagation] ($0.27 \geq \rho \geq 0.34$). Queues form, and in some cases platoons reach a stopped platoon, causing vehicles to shift platoons.
\item[Upstream propagation] ($0.35 \geq \rho \geq 0.4$). Queues can be long enough to reach an upstream intersection, potentially delaying vehicles more than one block upstream.
\end{description}

\item[Full capacity] ($0.41 \geq \rho \geq 0.6$). There is intermittent traffic flow but $J=J_{max}$, i.e.\ all intersections are utilized at their maximum capacity, with vehicles crossing constantly. Certainly, average velocities $\langle v\rangle$ decrease as density $\rho$ increases.

\item[Overutilized] ($0.6 \geq \rho \geq 0.79$). This subphase of the intermittent phase is characterized also by a flux $J<J_{max}$, but for reasons opposite to those for the underutilized subphase. At these high densities, rule 6 sometimes blocks all directions of an intersection, preventing any vehicle from crossing. This is not due to lack of vehicles, but to excess of vehicles. In fact, more than half of vehicles are stopped at any given moment. Thus, it is useful to focus on the movement of ``free space" between vehicles, which occurs in the opposite direction of traffic. Free spaces also group in ``space platoons". The dynamics of the space platoons are complementary to those of platoons in the underutilized intermittent subphase.  
In both subphases, flux $J$ is greater than maximum flux $J_{max}$ for triple intersections. The overutilized intermittent subphase can be further subdivided in two subphases for the double intersection scenario: downstream propagation and platoon propagation.
\begin{description}
\item[Downstream propagation]  ($0.6 \geq \rho \geq 0.67$). In this subphase, space platoons can be long enough to cover more than one intersection. In this way, a space platoon can sometimes propagate traffic flow downstream, so that vehicles may cross two intersections without stopping.
\item[Platoon propagation]  ($0.68 \geq \rho \geq 0.79$). Space platoons are shorter in this subphase, so vehicles need to stop at least once before crossing any intersection.
\end{description}

\end{description}

\item[Quasi-gridlock] ($0.8 \geq \rho \geq 0.94$). Almost all vehicles are stopped, but rule 6 prevents gridlock. Short space platoons are formed, and they rarely have to ``stop" flowing opposite to the vehicles' direction. This phase is complementary to the quasi-free flow phase.
This phase can be further subdivided in two subphases for the double intersection scenario: continuous flow and partial flow.

\begin{description}
\item[Continuous flow] ($0.8 \geq \rho \geq 0.88$). Traffic flows on all streets.
\item[Partial flow] ($0.89 \geq \rho \geq 0.94$). Traffic flows only on some streets, as extreme densities prevent flow on some roads. Considering that there are 12 intersections per street of 160 cells, a density $\rho>0.925$ for a particular street implies no flow $J=0$.
\end{description}

\item[Gridlock]  ($\rho > 0.94$). In this phase, all traffic is stopped ($J=v=0$). Due to initial conditions, intersections are blocked and traffic can flow on no street. This is complementary to the free flow phase, where no vehicle is stopped ($v=1$).

\end{description}

For the \emph{self-organizing} method, the triple intersection scenario offers \emph{full capacity} for almost half of the density spectrum. However, the maximum flux $J_{max}=1/6$ is 50\% less than for the double intersection scenario $J_{max}=1/4$. It is clear that, for most densities, the double intersection scenario---even when there are three times more intersections to be coordinated---offers more efficient traffic flow than the triple intersection scenario.

\subsection{Mixed scenario}

\begin{figure}[htbp]
\begin{center}
\includegraphics[width=.85\textwidth]{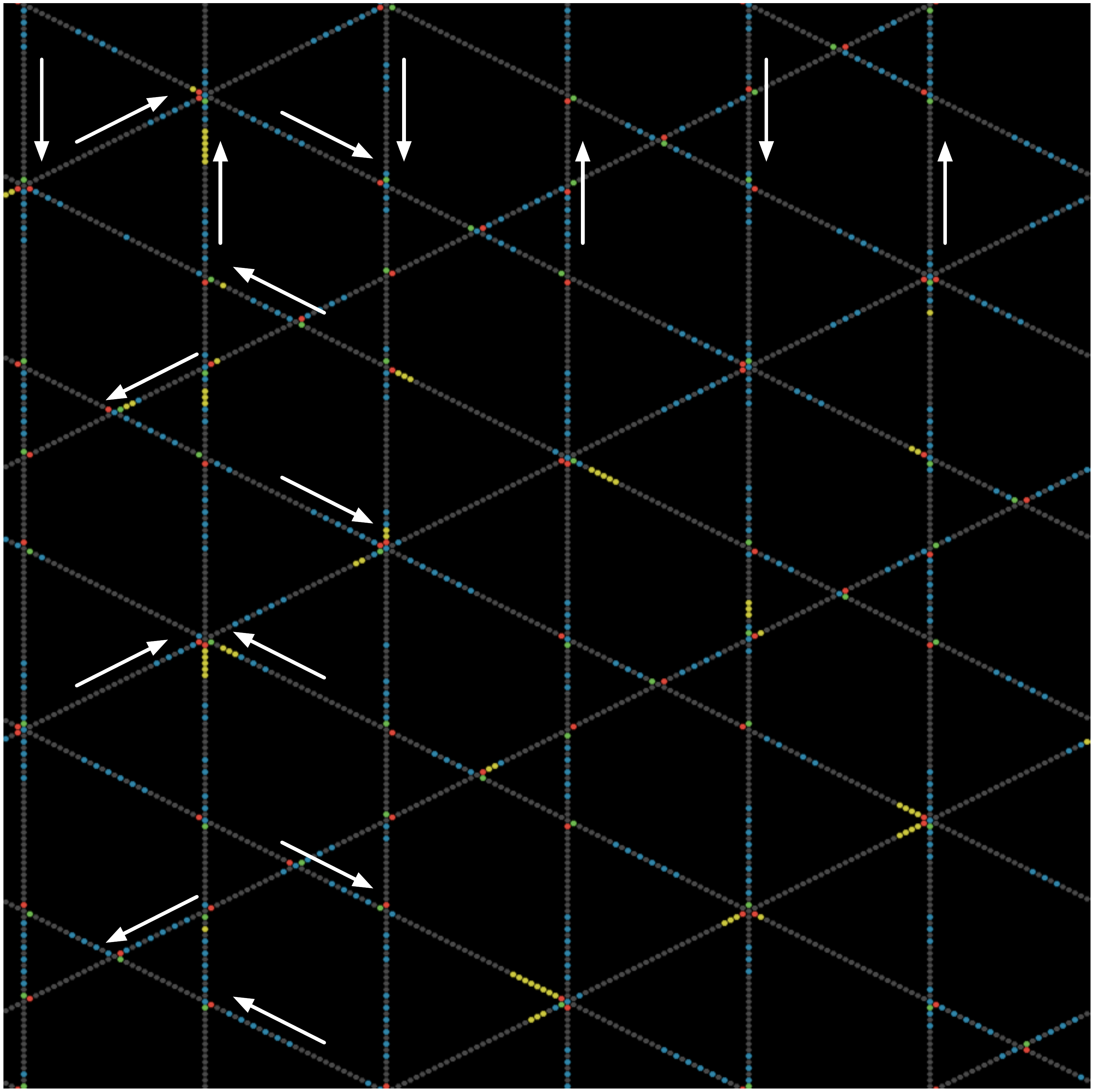}
\caption{Screenshot of mixed scenario with 16 streets and 12 triple and 48 double intersections, $\rho=1/5$.  Moving vehicles are colored with \textcolor{sky}{sky blue} and stopped vehicles are colored in \textcolor{yell}{yellow}.}
\label{fig:12triple48double}
\end{center}
\end{figure}

A third scenario is considered, where double and triple intersections are used. There are 16 streets, three streets in each of four directions and two streets in each of other two directions. Streets are arranged in such a way that there are 12 triple and 48 double intersections, as shown in Figure \ref{fig:12triple48double}. The triple intersections, as was already mentioned, have $50\%$ less flow capacity than double intersections. Of the 16 streets, there are only two with no triple intersections, while there are only two with only triple intersections, i.e.\ there are 12 streets with both intersection types.

\begin{figure}
     \centering
     \subfigure{
          \label{fig:results_city_mixA}
          \includegraphics[width=.55\textwidth]{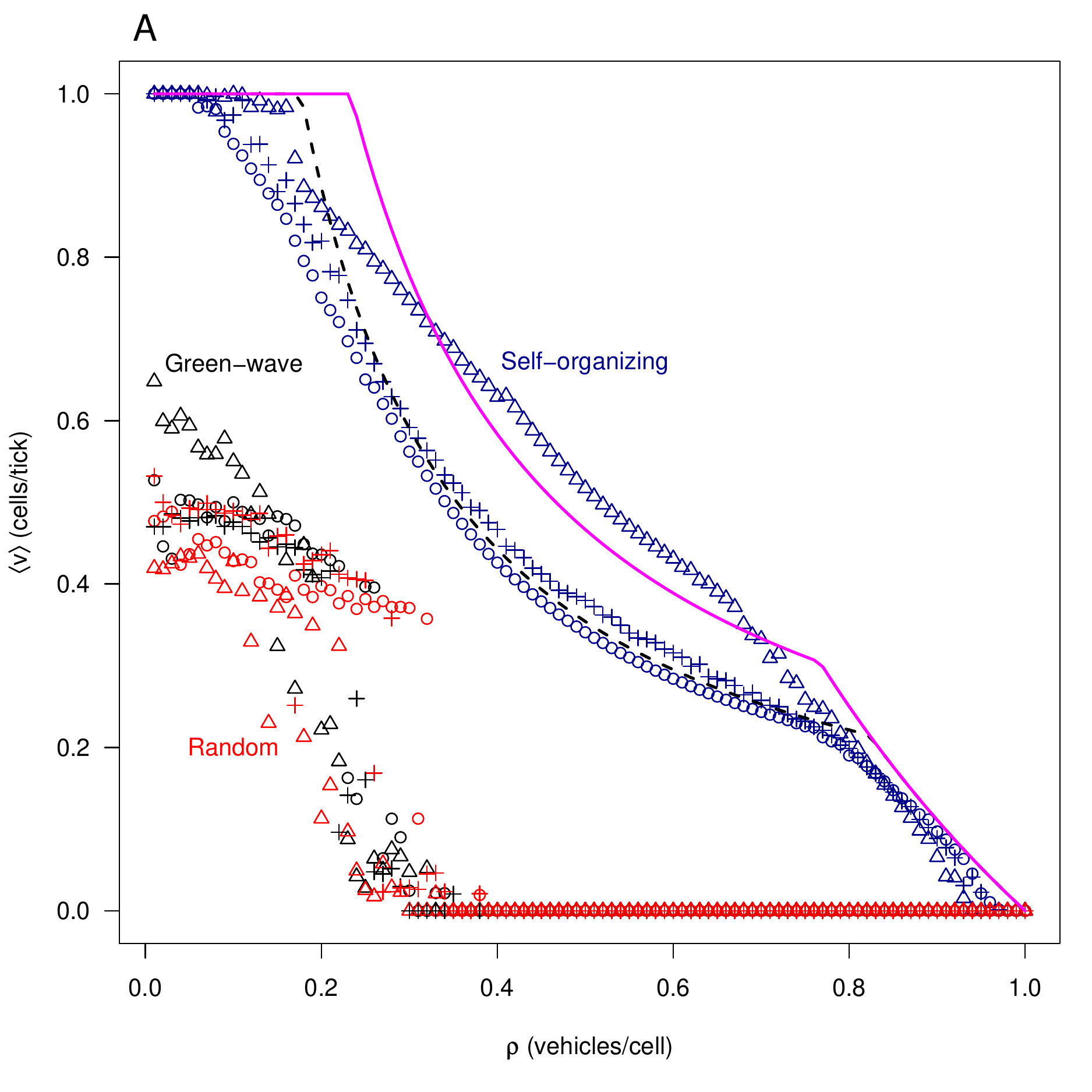}}
     \\     
     \subfigure{
          \label{fig:results_city_mixB}
          \includegraphics[width=.55\textwidth]{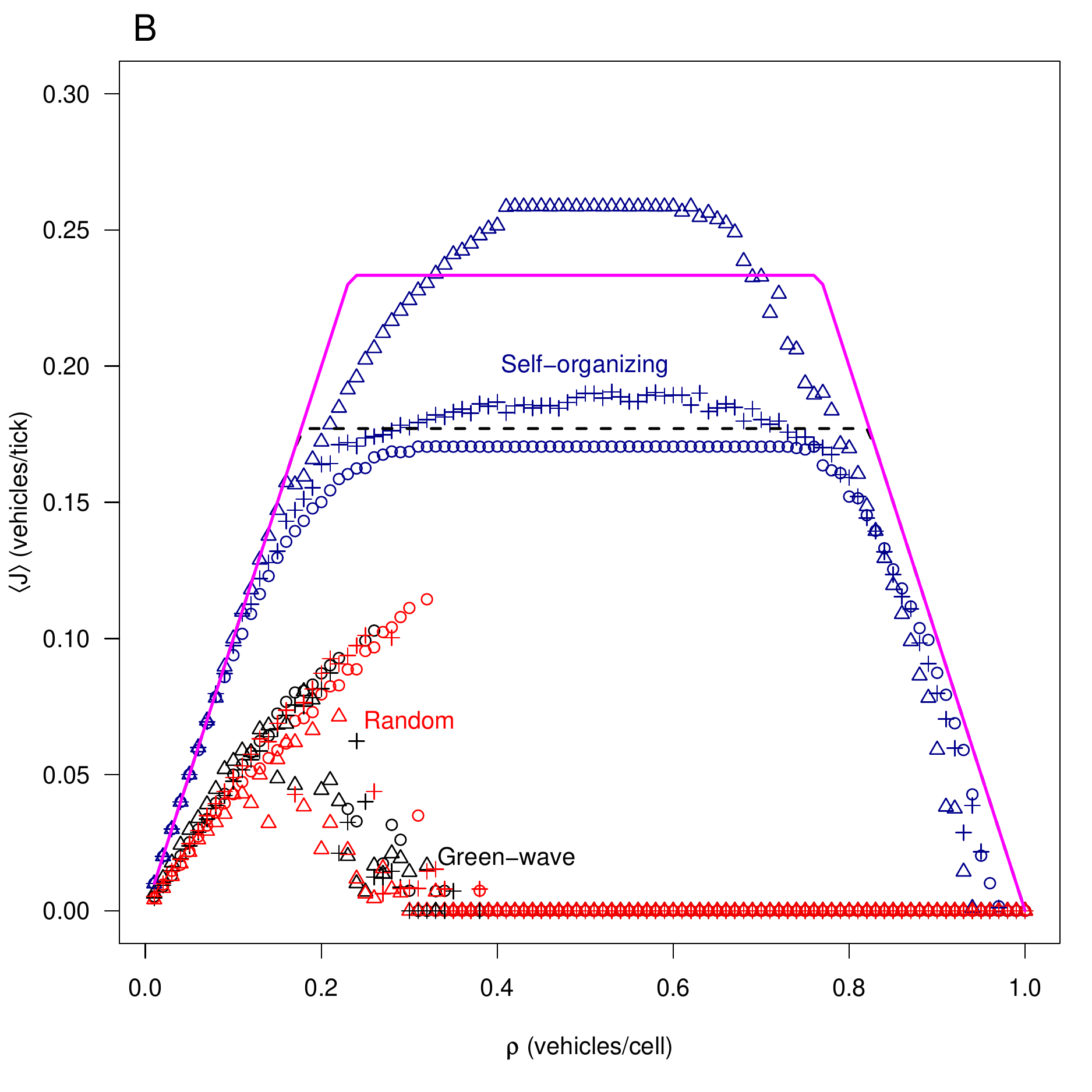}}

     \caption{Simulation results for 18 and 16 street scenarios: (A) average velocity $\langle v\rangle$ and (B) average flux $\langle J\rangle$ for different densities $\rho$, \emph{green-wave} (black), \emph{self-organizing} (\textcolor{blue}{blue}), and \emph{random} (\textcolor{red}{red}) methods. Three scenarios are considered: 36 triple intersections ({\Large $\circ$}), 108 double intersections ($\bigtriangleup$), and mixed ($+$, 12 triple with 48 double intersections). Dashed lines indicate optimality curves for mixed scenario when street constraints are considered ($J_{max}\approx0.177$), while \textcolor{magenta}{magenta} continous lines indicate optimality curves for mixed scenario when isolated intersections are considered ($J_{max}\approx0.233$).}
     \label{fig:results_city_mix}
\end{figure}

Intuitively, one could think that streets with at least one triple intersection would have their capacity limited by this intersection, i.e.\ $J_{max}=1/6$. If 14 out of 16 streets (all of same length) have at least one triple intersection, we could assume that the average maximum flux of the scenario would be $J_{max}=\frac{14\cdot \frac{1}{6}+2\cdot\frac{1}{4}}{16} = \frac{17}{96} \approx 0.177$. Still, the \emph{self-organizing method} achieves $J_{max} >  0.177$, especially for $\rho>1/2$ ($J_{max}\approx 0.19$). This implies that even in streets with some intersections with triple intersections, the traffic lights are coordinated in such a way that the average flow can be even higher than the expected one. This is because, even when the flux is limited by triple intersections, double intersections can take advantage of their higher capacity. If a triple intersection delays a double intersection upstream, the street crossing at the double intersection will take the opportunity to flow as much as possible. In this way, vehicles on streets with six double intersections and two triple intersections in practice will flow faster than vehicles on streets with six triple intersections. Certainly, vehicles on streets with twelve double intersections will flow even faster.
 The simulation results for this mixed scenario, together with the results of the previous two scenarios already discussed, are shown in Figure \ref{fig:results_city_mix}.

The \emph{green-wave} method performs even worse in the mixed scenario than in the triple scenario. The reason is that the fixed coordination of mixed traffic lights becomes even more complicated, and there is no free flow on any street. Since the duration of green phases is the same for all intersections, triple intersections have a period $50\%$ longer than double intersections. Again, the performance is comparable with that of the \emph{random} method.

The \emph{self-organizing} method, as could be expected, has a performance lying between the double and triple scenarios. However, as mentioned above, it reaches a maximum flux higher than the expected one, according to the number of streets with at least one triple intersection. There is free flow for low densities, showing that the \emph{self-organizing method} manages to coordinate traffic lights in such a complex scenario. Nevertheless, given the mixed topology, there is no full capacity intermittent phase, which would imply a flux $J=\frac{12\cdot \frac{1}{6}+48\cdot\frac{1}{4}}{60}=\frac{7}{30}\approx0.233$, i.e.\ all intersections being utilized at all times. If we take into account the criterion of isolated intersections, this yields a different set of optimality curves, where the \emph{self-organizing} method is farther from the optima, i.e. interference $\varPhi$ is higher. This is because different capacities of double and triple intersections create more interferences between intersections. This capacity difference makes it impossible to reach a state of maximum flux for all intersections, i.e. full capacity intermittent phase. Nevertheless, it should be noted that the system performance is better in the mixed scenario than in the scenario with only triple intersections. This implies that the \emph{self-organizing} method can take advantage of the increased capacity of some of the intersections, even if not optimally, in spite of the increased coordination complexity (more traffic lights in the mixed scenario (60) than in the triple scenario (36)).

\subsection{Optimality and Interference}

\begin{figure}
     \centering
     \subfigure{
          \label{fig:PhiA}
          \includegraphics[width=.55\textwidth]{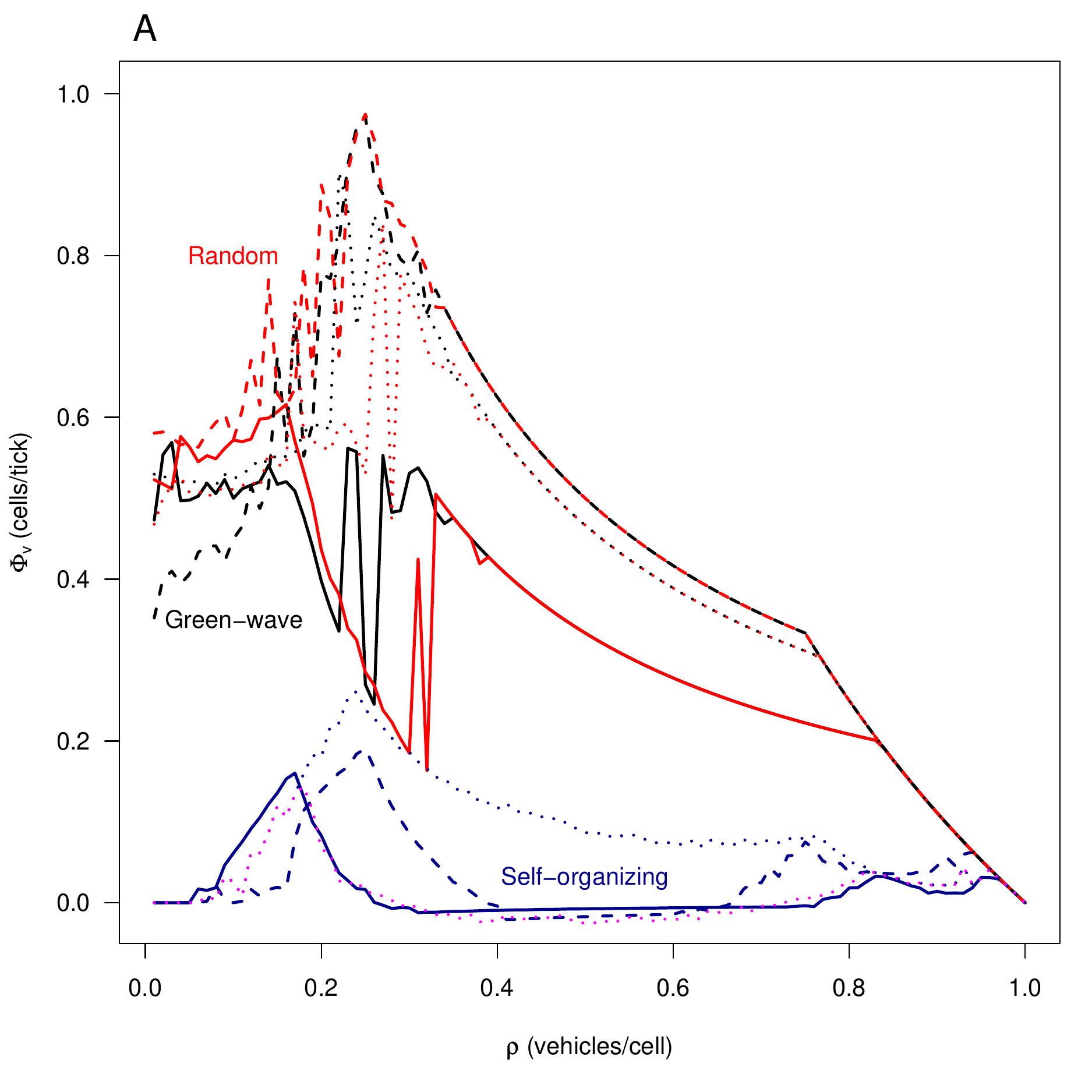}}
     \\     
     \subfigure{
          \label{fig:PhiB}
          \includegraphics[width=.55\textwidth]{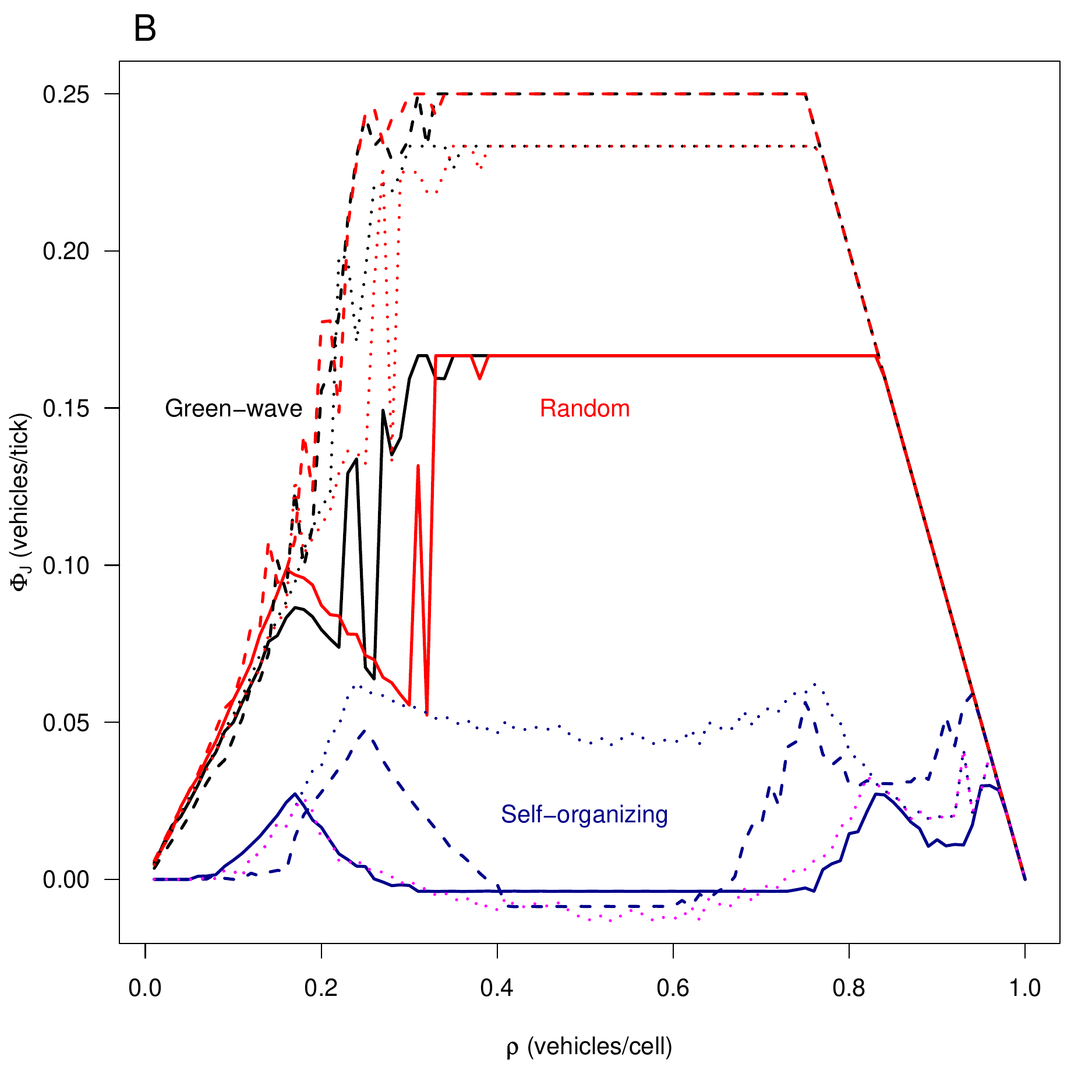}}

     \caption{Interference curves for triple (continuous lines), double (dashed lines) and mixed (pointed lines) scenarios: (A) velocity interference $\varPhi_v$ and (B) flux interference $\varPhi_J$  for different densities $\rho$, \emph{green-wave} (black), \emph{self-organizing} (\textcolor{blue}{blue}), and \emph{random} (\textcolor{red}{red}) methods. \textcolor{magenta}{magenta} pointed lines indicate interference curves assuming a $J_{max}\approx0.177$ in the mixed scenario.}
     \label{fig:Phi}
\end{figure}

We can calculate how close to optimality different methods are comparing the optimality curves for isolated intersections with the experimental results for different scenarios. This difference was defined as interference $\varPhi$ in equations \ref{eq:Phiv} and \ref{eq:PhiJ}. The interference curves for different methods are shown in Figure \ref{fig:Phi}, while the values of their integrals are shown in Tables \ref{table:Phiv} and \ref{table:PhiJ}.

For all methods, the higher velocity interference $\varPhi_{v}$ can be found near $\rho=J_{max}$. The reason is that this is the region where theoretically there is the highest possible velocity (free flow) for the highest possible density ($J_{max}$). This is the region where more improvements can be made concerning velocity in the problem of coordinating traffic lights.

Since the \emph{green-wave} and \emph{random} methods reach a gridlock for relatively low densities, the flux interference $\varPhi_{J}$ is highest when the flux should be highest, i.e. $J=J_{max}$.
For the \emph{self-organizing} method, there is a higher flux interference $\varPhi_{J}$ within the underutilized intermittent and overutilized intermittent phases. There is low interference within the free-flow, quasi-free flow, and full capacity intermittent phases. In these regions of the density space, the \emph{self-organizing method} is optimal or close to optimal. There is certain interference in the quasi-gridlock, partial flow and gridlock phases, since theoretically traffic still should flow in an isolated intersection, while in a city scenario initial conditions block the flow of some or all of the streets.

The large spikes near some phase transitions are due to single runs per density value. These are smoothened with a larger statistical sample, which has high standard deviations precisely near the same phase transitions where the spikes are present.

\begin{table}[htdp]
\caption{Velocity interferences $\varPhi_v$ for different methods and scenarios. \textcolor{magenta}{Magenta} row assumes a $J_{max}\approx0.177$ in the mixed scenario.}
\begin{center}
\begin{tabular}{|c|c|c|c|c|c|c|}
\hline
scenario	&\textcolor{blue}{$\varPhi_{v}$ Self-organizing}	&$\varPhi_{v}$ Green-wave	&\textcolor{red}{$\varPhi_{v}$ Random}\\
\hline
triple	&\textcolor{blue}{0.01543474}	&0.3274843	&\textcolor{red}{0.3174297}	\\
\hline
double	&\textcolor{blue}{0.03256081}	&0.4484157	&\textcolor{red}{0.4759376}	\\
\hline
mixed, $J_{max}\approx0.233$	&\textcolor{blue}{0.08689782}	& 0.4336379	&\textcolor{red}{0.416929}	\\

\textcolor{magenta}{mixed, $J_{max}\approx0.177$}	&\textcolor{blue}{0.01056809}	&	&\textcolor{red}{}	\\

\hline

\end{tabular}
\end{center}
\label{table:Phiv}
\end{table}%

\begin{table}[htdp]
\caption{Flux interferences $\varPhi_J$ for different methods and scenarios.
\textcolor{magenta}{Magenta} row assumes a $J_{max}\approx0.177$ in the mixed scenario.}
\begin{center}
\begin{tabular}{|c|c|c|c|c|c|c|}
\hline
scenario	&\textcolor{blue}{$\varPhi_{J}$ Self-organizing}	&$\varPhi_{J}$ Green-wave	&\textcolor{red}{$\varPhi_{J}$ Random}\\
\hline
triple	&\textcolor{blue}{0.004418822}	& 0.1234725	&\textcolor{red}{0.1190469}	\\
\hline
double	&\textcolor{blue}{0.01471438}	& 0.1759046	&\textcolor{red}{0.1786202}	\\
\hline
mixed, $J_{max}\approx0.233$	&\textcolor{blue}{0.03700456}	& 0.1664656	&\textcolor{red}{0.1625401}	\\
\textcolor{magenta}{mixed, $J_{max}\approx0.177$}	&\textcolor{blue}{0.003842065}	&	&\textcolor{red}{}	\\

\hline

\end{tabular}
\end{center}
\label{table:PhiJ}
\end{table}%

The interferences $\varPhi_v$ and $\varPhi_J$ (equations \ref{eq:Phiv} and \ref{eq:PhiJ}), shown in Tables \ref{table:Phiv} and \ref{table:PhiJ}, are similar for the \emph{green-wave} and \emph{random} methods. The \emph{green-wave} method is more efficient for lower densities and for the double scenario, but worse than \emph{random} for medium densities (it reaches gridlock earlier) and for the triple and mixed scenarios. The difference of values for the interferences for different scenarios is related more to differences of $J_{max}$ than to differences of performance, since there is maximum interference ($v=J=0$, i.e. gridlock) for a broad range of densities.

The interferences $\varPhi_v$ and $\varPhi_J$ for the \emph{self-organizing} method are much lower in all scenarios. In the triple scenario, it is close to theoretical optimality, which is achieved when $\varPhi_J=0$. Still, the increased capacity of the double scenario gives a better performance, even when the interference is higher. This is also the case for the mixed scenario, where the interference is highest because the lack of a full-capacity intermittent phase.

\section{Discussion}
\label{sec:discussion}

The experiments showed that the \emph{self-organizing} method is highly scalable. It manages to coordinate flow in at least six different directions and prevent gridlocks. This suggests that for the traffic light coordination problem, an adaptive mechanism, in this case achieved with self-organization, will be remarkably more efficient than a fixed prediction-based mechanism. This is because the traffic coordination problem is non-stationary~\citep{GershensonDCSOS}. The positions of vehicles are changing constantly and are interdependent. The optimal coordination of traffic lights changes with each vehicular configuration, which cannot be predicted.
Many adaptive methods in the literature~\citep{FHA2005,Henry:1983,Mauro:1990,Robertson:1991,FaietaHuberman1993,Gartner:2001,Diakaki:2003,FouladvandEtAl2004a,Mirchandani:2005,Bazzan2005} adjust parameters at a timescale considerably larger than the one in which changes in the traffic configuration occur. Certainly, the adaptation will offer some improvements over static methods. However, methods that make adjustments at the seconds scale will offer a better performance, since the predictable horizon of city traffic is limited to a couple of minutes~\citep{HelbingEtAl2005,Gershenson2005,CoolsEtAl2007,Lammer:2008}.
An adaptive method at short timescales, as the one presented here, will be able to adjust and regulate the traffic conditions with considerable improvements over current fixed-cycle methods, such as the \emph{green-wave} used here.

A clear insight from our simulations for city planning is that complex intersections should be avoided, since these function as bottlenecks. This was already known~\citep{Buchanon:1963,Bach:2006}. However, an interesting result presented here  is that without proper light coordination, the benefit of an efficient topology is almost lost, and can even be counterproductive. This can be seen with the similar performances of the \emph{green-wave} method in different scenarios. A fixed-cycle method cannot take advantage of an efficient topology. This is important especially for cities where there are already complex intersections and the street topology can be difficult to modify.

Building on our previous results~\citep{Gershenson2005,CoolsEtAl2007,GershensonRosenblueth:2010}, the simulations presented here showed that the \emph{self-organizing} method is highly scalable, and has a graceful degradation of performance as density increases. Moreover, this scalability enables the method to exploit much better the city topology.

An important issue that we have not considered in the simulations
presented here is that of turning vehicles. 
In principle, turning vehicles do not limit the \emph{self-organizing} method, as shown in
previous work~\citep{Gershenson2005,CoolsEtAl2007}. However, if a
vehicle is turning on a blocked street, this blockage could propagate
to crossing streets. To prevent this, dedicated turning lanes should
be used, where a traffic light for turning vehicles will be set to red
if the crossing street has a blockage ahead. This would be a simple
extension of rule 5. 

Another issue to be considered before implementing the \emph{self-organizing} method in a city is that of pedestrians. Preliminary results considering pedestrians have been encouraging, and will be presented in future work.

\section{Conclusions}
\label{sec:conclusions}

By definition, models are caricatures of the world.
The objective of such caricatures is to abstract away details so as
to foil properties of interest.
Hence, a model should be as simple as possible, provided such
properties are highlighted.
Different models will therefore be useful for different purposes.

In the case of vehicular traffic in intersections
controlled with traffic lights, the more complex and therefore
more realistic models can reproduce numerous details.
An example is our own model based on agents~\citep{Gershenson2005, CoolsEtAl2007,deGier:2011}.
Such a model does exhibit a behavior reproducing many features.
However, if we are interested in identifying phases, this agent-based
model is not useful.
The reason is that the phase transitions are smoothed out.

We have consequently considered a simpler model based on cellular automata.
Such automata are valuable because of
their simplicity and for nonetheless exhibiting a similar behavior to
that of more elaborated models.
Moreover, our model employs only elementary automata, where the state
of a cell in the next tick is a function only of the current state of
such a cell and those of its nearest neighbors.
This model is computationally cheap allowing the simulations of
numerous vehicles.
Furthermore, this model is superior to our previous mult-agent one, as
the phase transitions are conspicuous.

Another advantage of our simple model of city traffic is that optimality can be deduced in a straightforward fashion, enabling the comparison of methods for coordinating traffic lights with a theoretical optimum. This informs which regions of the density space have an already optimal behavior and which regions can still be improved, at least in theory. Moreover, we were able to show that our proposed self-organizing method is close to optimality, independently of the poor performance of the green-wave method. This is relevant, because there were no previous benchmarks for comparing the performance of different traffic light controllers.

The proposed optimality measurement---where the performance of a traffic light controller is compared against the capacity of an isolated single intersection---can be also used with other models of city traffic for benchmarking. If the optimality curves (equations \ref{eq:voptim} and \ref{eq:Joptim}) cannot be deduced analytically, experimental data can be fitted and used to calculate interference $\varPhi$.

\section*{Acknowledgments}

We gratefully acknowledge the facilities provided by IIMAS, UNAM.
C.G. was partially supported by SNI membership 47907 of CONACyT, Mexico. 


\bibliographystyle{cgg}
\bibliography{carlos,traffic,sos,complex,RBN}

\begin{thebibliography}{}

\bibitem[\protect\citeauthoryear{Ball}{Ball}{2004}]{Ball:2004}
{\sc Ball, P.} (2004).
\newblock Beating the lights.
\newblock {\em News@Nature\/}.
\newblock URL \url{http://tinyurl.com/lsqopz}.

\bibitem[\protect\citeauthoryear{Bazzan}{Bazzan}{2005}]{Bazzan2005}
{\sc Bazzan, A. L.~C.} (2005).
\newblock A distributed approach for coordination of traffic signal agents.
\newblock {\em Autonomous Agents and Multiagent Systems\/}~{\bf 10\/}~(1)
  (March): 131--164.
\newblock URL \url{http://tinyurl.com/2vld8y}.

\bibitem[\protect\citeauthoryear{Biham, Middleton, and Levine}{Biham
  et~al\mbox{.}}{1992}]{biham:middleton:levine:92}
{\sc Biham, O.}, {\sc Middleton, A.~A.}, {\sc and} {\sc Levine, D.} (1992).
\newblock Self organization and a dynamical transition in traffic flow models.
\newblock {\em Physical Review A\/}~{\bf 46\/}~(10): R6124--R6217.
\newblock URL \url{http://arxiv.org/abs/cond-mat/9206001}.

\bibitem[\protect\citeauthoryear{Brockfeld, Barlovic, Schadschneider, and
  Schreckenberg}{Brockfeld et~al\mbox{.}}{2001}]{BrockfeldEtAl2001}
{\sc Brockfeld, E.}, {\sc Barlovic, R.}, {\sc Schadschneider, A.}, {\sc and}
  {\sc Schreckenberg, M.} (2001).
\newblock Optimizing traffic lights in a cellular automaton model for city
  traffic.
\newblock {\em Physical Review E\/}~{\bf 64}: 056132.
\newblock URL \url{http://dx.doi.org/10.1103/PhysRevE.64.056132}.

\bibitem[\protect\citeauthoryear{Buchanon}{Buchanon}{1963}]{Buchanon:1963}
{\sc Buchanon, C.} (1963).
\newblock {\em Traffic in Towns; A study of long term problems of traffic in
  urban areas}.
\newblock HMSO, London.

\bibitem[\protect\citeauthoryear{Chopard, Luthi, and Queloz}{Chopard
  et~al\mbox{.}}{1996}]{chopard:luthi:queloz:96}
{\sc Chopard, B.}, {\sc Luthi, P.~O.}, {\sc and} {\sc Queloz, P.-A.} (1996).
\newblock Cellular automata model of car traffic in a two-dimensional street
  network.
\newblock {\em Journal of Physics A: Mathematics and General\/}~{\bf 29}:
  2325--2336.
\newblock URL \url{http://dx.doi.org/10.1088/0305-4470/29/10/012}.

\bibitem[\protect\citeauthoryear{Chowdhury, Santen, and
  Schadschneider}{Chowdhury et~al\mbox{.}}{2000}]{ChowdhuryEtAl2000}
{\sc Chowdhury, D.}, {\sc Santen, L.}, {\sc and} {\sc Schadschneider, A.}
  (2000).
\newblock Statistical physics of vehicular traffic and some related systems.
\newblock {\em Physics Reports\/}~{\bf 329\/}~(4-6): 199 -- 329.
\newblock URL \url{http://dx.doi.org/10.1016/S0370-1573(99)00117-9}.

\bibitem[\protect\citeauthoryear{Chowdhury and Schadschneider}{Chowdhury and
  Schadschneider}{1999}]{chowdhury:schadschneider:99}
{\sc Chowdhury, D.} {\sc and} {\sc Schadschneider, A.} (1999).
\newblock Self-organization of traffic jams in cities: Effects of stochastic
  dynamics and signal periods.
\newblock {\em Physical Review E\/}~{\bf 59\/}~(2): R1311--R1314.
\newblock URL \url{http://arxiv.org/abs/cond-mat/9810099}.

\bibitem[\protect\citeauthoryear{Cools, Gershenson, and {D'Hooghe}}{Cools
  et~al\mbox{.}}{2007}]{CoolsEtAl2007}
{\sc Cools, S.~B.}, {\sc Gershenson, C.}, {\sc and} {\sc {D'Hooghe}, B.}
  (2007).
\newblock Self-organizing traffic lights: A realistic simulation.
\newblock In {\em Self-Organization: Applied Multi-Agent Systems},
  {M.~Prokopenko}, (Ed.). Springer, Chapter~3, pp. 41--49.
\newblock URL \url{http://uk.arxiv.org/abs/nlin.AO/0610040}.

\bibitem[\protect\citeauthoryear{Cremer and Ludwig}{Cremer and
  Ludwig}{1986}]{cremer:ludwig:86}
{\sc Cremer, M.} {\sc and} {\sc Ludwig, J.} (1986).
\newblock A fast simulation model for traffic flow on the basis of {B}oolean
  operations.
\newblock {\em Mathematics and Computers in Simulation\/}~{\bf 28}: 297--303.

\bibitem[\protect\citeauthoryear{{de Gier}, Garoni, and Rojas}{{de Gier}
  et~al\mbox{.}}{2011}]{deGier:2011}
{\sc {de Gier}, J.}, {\sc Garoni, T.~M.}, {\sc and} {\sc Rojas, O.} (2011).
\newblock Traffic flow on realistic road networks with adaptive traffic lights.
\newblock arXiv 1011.6211.
\newblock URL \url{http://arxiv.org/abs/1011.6211}.

\bibitem[\protect\citeauthoryear{{de Jong}, {van Hal}, and {de Jong}}{{de Jong}
  et~al\mbox{.}}{2006}]{Bach:2006}
{\sc {de Jong}, T.}, {\sc {van Hal}, E.}, {\sc and} {\sc {de Jong}, M.}, Eds.
  (2006).
\newblock {\em Urban design and traffic: a selection from {Bach's} toolbox}.
\newblock CROW, Ede, Netherlands.

\bibitem[\protect\citeauthoryear{Diakaki, Dinopoulou, Aboudolas, Papageorgiou,
  Ben-Shabat, Seider, and Leibov}{Diakaki et~al\mbox{.}}{2003}]{Diakaki:2003}
{\sc Diakaki, C.}, {\sc Dinopoulou, V.}, {\sc Aboudolas, K.}, {\sc
  Papageorgiou, M.}, {\sc Ben-Shabat, E.}, {\sc Seider, E.}, {\sc and} {\sc
  Leibov, A.} (2003).
\newblock Extensions and new applications of the traffic signal control
  strategy tuc.
\newblock {\em Transportation Research Record\/}~{\bf 1856}: 202--211.
\newblock URL \url{http://www.aimsun.com/site/content/view/66/35/}.

\bibitem[\protect\citeauthoryear{Faieta and Huberman}{Faieta and
  Huberman}{1993}]{FaietaHuberman1993}
{\sc Faieta, B.} {\sc and} {\sc Huberman, B.~A.} (1993).
\newblock Firefly: A synchronization strategy for urban traffic control.
\newblock Tech. Rep. SSL-42, Xerox PARC, Palo Alto, CA.

\bibitem[\protect\citeauthoryear{{Federal Highway Administration}}{{Federal
  Highway Administration}}{2005}]{FHA2005}
{\sc {Federal Highway Administration}}. (2005).
\newblock {\em Traffic Control Systems Handbook}.
\newblock U.S. Department of Transportation.
\newblock URL \url{http://tinyurl.com/4zsb4r}.

\bibitem[\protect\citeauthoryear{Fouladvand, Sadjadi, and Shaebani}{Fouladvand
  et~al\mbox{.}}{2004}]{FouladvandEtAl2004a}
{\sc Fouladvand, M.~E.}, {\sc Sadjadi, Z.}, {\sc and} {\sc Shaebani, M.~R.}
  (2004).
\newblock Optimized traffic flow at a single intersection: Traffic responsive
  signalization.
\newblock {\em J. Phys. A: Math. Gen.\/}~{\bf 37}: 561--576.
\newblock URL \url{http://dx.doi.org/10.1088/0305-4470/37/3/002}.

\bibitem[\protect\citeauthoryear{Fukui and Ishibashi}{Fukui and
  Ishibashi}{1996}]{fukui:ishibashi:96}
{\sc Fukui, M.} {\sc and} {\sc Ishibashi, Y.} (1996).
\newblock Traffic flow in 1{D} cellular automaton model including cars moving
  with high speed.
\newblock {\em Journal of the Physical Society of Japan\/}~{\bf 65\/}~(6):
  1868--1870.
\newblock URL \url{http://dx.doi.org/10.1143/JPSJ.65.1868}.

\bibitem[\protect\citeauthoryear{Gartner, Little, and Gabbay}{Gartner
  et~al\mbox{.}}{1975}]{N.H.Gartner:1975}
{\sc Gartner, N.}, {\sc Little, J.}, {\sc and} {\sc Gabbay, H.} (1975).
\newblock Optimization of traffic signal settings by mixed integer linear
  programming.
\newblock {\em Transportation Science\/}~{\bf 9}: 321--363.
\newblock URL \url{http://ntlsearch.bts.gov/tris/record/tris/00131180.html}.

\bibitem[\protect\citeauthoryear{Gartner, Pooran, and Andrews}{Gartner
  et~al\mbox{.}}{2001}]{Gartner:2001}
{\sc Gartner, N.~H.}, {\sc Pooran, F.~J.}, {\sc and} {\sc Andrews, C.~M.}
  (2001).
\newblock Implementation of the {OPAC} adaptive control strategy in a trafffic
  signaling network.
\newblock In {\em IEEE Intelligent Transportation Systems Conference
  Proceedings}. pp.~195--200.

\bibitem[\protect\citeauthoryear{Gershenson}{Gershenson}{2005}]{Gershenson2005}
{\sc Gershenson, C.} (2005).
\newblock Self-organizing traffic lights.
\newblock {\em Complex Systems\/}~{\bf 16\/}~(1): 29--53.
\newblock URL \url{http://www.complex-systems.com/pdf/16-1-2.pdf}.

\bibitem[\protect\citeauthoryear{Gershenson}{Gershenson}{2007}]{GershensonDCSOS}
{\sc Gershenson, C.} (2007).
\newblock {\em Design and Control of Self-organizing Systems}.
\newblock CopIt Arxives, Mexico.
\newblock http://tinyurl.com/DCSOS2007.
\newblock URL \url{http://tinyurl.com/DCSOS2007}.

\bibitem[\protect\citeauthoryear{Gershenson}{Gershenson}{2011}]{Gershenson:2010a}
{\sc Gershenson, C.} (2011).
\newblock The sigma profile: A formal tool to study organization and its
  evolution at multiple scales.
\newblock {\em Complexity\/}~{\bf {In Press}}.
\newblock URL \url{http://arxiv.org/abs/0809.0504}.

\bibitem[\protect\citeauthoryear{Gershenson and Rosenblueth}{Gershenson and
  Rosenblueth}{2010}]{GershensonRosenblueth:2010}
{\sc Gershenson, C.} {\sc and} {\sc Rosenblueth, D.~A.} (2010).
\newblock Adaptive self-organization vs. static optimization: A qualitative
  comparison in traffic light coordination.
\newblock Tech. Rep. 2010.03, C3.

\bibitem[\protect\citeauthoryear{Helbing, L\"{a}mmer, and Lebacque}{Helbing
  et~al\mbox{.}}{2005}]{HelbingEtAl2005}
{\sc Helbing, D.}, {\sc L\"{a}mmer, S.}, {\sc and} {\sc Lebacque, J.-P.}
  (2005).
\newblock Self-organized control of irregular or perturbed network traffic.
\newblock In {\em Optimal Control and Dynamic Games}, {C.~Deissenberg} {and}
  {R.~F. Hartl}, (Eds.). Springer, Dordrecht, 239--274.
\newblock URL \url{http://arxiv.org/abs/physics/0511018}.

\bibitem[\protect\citeauthoryear{Henry, Farges, and Tuffal}{Henry
  et~al\mbox{.}}{1983}]{Henry:1983}
{\sc Henry, J.}, {\sc Farges, J.}, {\sc and} {\sc Tuffal, J.} (1983).
\newblock The {PRODYN} real time traffic algorithm.
\newblock In {\em Proceedings of the Int. Fed. of Aut. Control (IFAC) Conf}.
  Baden-Baden.

\bibitem[\protect\citeauthoryear{Kanai}{Kanai}{2010}]{Kanai:2010}
{\sc Kanai, M.} (2010).
\newblock Calibration of the particle density in cellular-automaton models for
  traffic flow.
\newblock {\em J. Phys. Soc. Jpn.\/}~{\bf 79}: 075002.
\newblock URL \url{http://arxiv.org/abs/1002.1382}.

\bibitem[\protect\citeauthoryear{L\"{a}mmer and Helbing}{L\"{a}mmer and
  Helbing}{2008}]{Lammer:2008}
{\sc L\"{a}mmer, S.} {\sc and} {\sc Helbing, D.} (2008).
\newblock Self-control of traffic lights and vehicle flows in urban road
  networks.
\newblock {\em J. Stat. Mech.\/}~{\bf P04019}.
\newblock URL \url{http://dx.doi.org/10.1088/1742-5468/2008/04/P04019}.

\bibitem[\protect\citeauthoryear{Maerivoet and {De Moor}}{Maerivoet and {De
  Moor}}{2005}]{Maerivoet:2005}
{\sc Maerivoet, S.} {\sc and} {\sc {De Moor}, B.} (2005).
\newblock Cellular automata models of road traffic.
\newblock {\em Physics Reports\/}~{\bf 419\/}~(1) (November): 1--64.
\newblock URL \url{http://arxiv.org/abs/physics/0509082}.

\bibitem[\protect\citeauthoryear{Mauro and {Di Taranto}}{Mauro and {Di
  Taranto}}{1990}]{Mauro:1990}
{\sc Mauro, V.} {\sc and} {\sc {Di Taranto}, D.} (1990).
\newblock {UTOPIA}.
\newblock In {\em Proceedings of the 6th IFAC / IFIP / IFORS Symposium on
  Control Computers and Communication in Transportation}. Paris, France.

\bibitem[\protect\citeauthoryear{Mirchandani and Wang}{Mirchandani and
  Wang}{2005}]{Mirchandani:2005}
{\sc Mirchandani, P.} {\sc and} {\sc Wang, F.-Y.} (2005).
\newblock {RHODES} to intelligent transportation systems.
\newblock {\em IEEE Intelligent Systems\/}~{\bf 20\/}~(1) (January--February):
  10--15.
\newblock URL \url{http://dx.doi.org/10.1109/MIS.2005.15}.

\bibitem[\protect\citeauthoryear{Moreira}{Moreira}{2003}]{Moreira:2003}
{\sc Moreira, A.} (2003).
\newblock Universality and decidability of number-conserving cellular automata.
\newblock {\em Theoretical Computer Science\/}~{\bf 292\/}~(3) (January):
  711--721.
\newblock URL \url{http://dx.doi.org/10.1016/S0304-3975(02)00065-8}.

\bibitem[\protect\citeauthoryear{Nagel and Paczuski}{Nagel and
  Paczuski}{1995}]{nagel:paczuski:95}
{\sc Nagel, K.} {\sc and} {\sc Paczuski, M.} (1995).
\newblock Emergent traffic jams.
\newblock {\em Physical Review E\/}~{\bf 51\/}~(4): 2909--2918.
\newblock URL \url{http://dx.doi.org/10.1103/PhysRevE.51.2909}.

\bibitem[\protect\citeauthoryear{Nagel and Schreckenberg}{Nagel and
  Schreckenberg}{1992}]{NaSch1992}
{\sc Nagel, K.} {\sc and} {\sc Schreckenberg, M.} (1992).
\newblock A cellular automaton model for freeway traffic.
\newblock {\em Journal of Physics I France\/}~{\bf 2}: 2221--2229.
\newblock URL \url{http://dx.doi.org/10.1051/jp1:1992277}.

\bibitem[\protect\citeauthoryear{Papadimitriou and Tsitsiklis}{Papadimitriou
  and Tsitsiklis}{1999}]{PapadimitriouTsitsiklis1999}
{\sc Papadimitriou, C.~H.} {\sc and} {\sc Tsitsiklis, J.~N.} (1999).
\newblock The complexity of optimal queuing network control.
\newblock {\em Mathematics of Operations Research\/}~{\bf 24\/}~(2) (February):
  293--305.
\newblock URL \url{http://portal.acm.org/citation.cfm?id=316389.316391}.

\bibitem[\protect\citeauthoryear{Robertson}{Robertson}{1969}]{Robertson:1969}
{\sc Robertson, D.} (1969).
\newblock Transyt: a traffic network study tool.
\newblock LR 253, Road Res. Lab., London.

\bibitem[\protect\citeauthoryear{Robertson and Bretherton}{Robertson and
  Bretherton}{1991}]{Robertson:1991}
{\sc Robertson, D.} {\sc and} {\sc Bretherton, R.} (1991).
\newblock Optimizing networks of traffic signals in real time---the {SCOOT}
  method.
\newblock {\em Vehicular Technology, IEEE Transactions on\/}~{\bf 40\/}~(1)
  (February): 11--15.

\bibitem[\protect\citeauthoryear{Rosenblueth and Gershenson}{Rosenblueth and
  Gershenson}{2011}]{RosenbluethGershenson:2010}
{\sc Rosenblueth, D.~A.} {\sc and} {\sc Gershenson, C.} (2011).
\newblock A model of city traffic based on elementary cellular automata.
\newblock {\em Complex Systems\/}~{\bf {In Press}}.

\bibitem[\protect\citeauthoryear{Schadschneider, Chowdhury, Brockfeld, Klauck,
  Santen, and Zittartz}{Schadschneider
  et~al\mbox{.}}{1999}]{schadschneider&:99}
{\sc Schadschneider, A.}, {\sc Chowdhury, D.}, {\sc Brockfeld, E.}, {\sc
  Klauck, K.}, {\sc Santen, L.}, {\sc and} {\sc Zittartz, J.} (1999).
\newblock A new cellular automata model for city traffic.
\newblock In {\em Proc. Traffic and Granular Flow '99: Social, Traffic, and
  Granular Dynamics}, {D.~Helbing}, {H.~J. Herrmann}, {M.~Schreckenberg}, {and}
  {D.~E. Wolf}, (Eds.).
\newblock URL \url{http://arxiv.org/abs/cond-mat/9911312}.

\bibitem[\protect\citeauthoryear{Simon and Nagel}{Simon and
  Nagel}{1997}]{simon:nagel:97}
{\sc Simon, P.~M.} {\sc and} {\sc Nagel, K.} (1997).
\newblock A simplified cellular automaton model for city traffic.
\newblock Tech. Rep. LA-UR 97-707, LANL.
\newblock URL \url{http://arxiv.org/abs/cond-mat/9801022v1}.

\bibitem[\protect\citeauthoryear{Sims and Dobinson}{Sims and
  Dobinson}{1980}]{SCATS1980}
{\sc Sims, A.~G.} {\sc and} {\sc Dobinson, K.~W.} (1980).
\newblock The {Sydney} coordinated adaptive traffic ({SCAT}) system, philosophy
  and benefits.
\newblock {\em IEEE Transactions of Vehicular Technology\/}~{\bf VT-29\/}~(2)
  (May): 130--137.

\bibitem[\protect\citeauthoryear{T{\"o}r{\"o}k and Kert{\'e}sz}{T{\"o}r{\"o}k
  and Kert{\'e}sz}{1996}]{TorokKertesz1999}
{\sc T{\"o}r{\"o}k, J.} {\sc and} {\sc Kert{\'e}sz, J.} (1996).
\newblock The green wave model of two-dimensional traffic: Transitions in the
  flow properties and in the geometry of the traffic jam.
\newblock {\em Physica A\/}~{\bf 231\/}~(4) (October): 515--533.
\newblock URL \url{http://dx.doi.org/10.1016/0378-4371(96)00144-6}.

\bibitem[\protect\citeauthoryear{Wilensky}{Wilensky}{1999}]{Wilensky1999}
{\sc Wilensky, U.} (1999).
\newblock {NetLogo}.
\newblock URL \url{http://ccl.northwestern.edu/netlogo}.

\bibitem[\protect\citeauthoryear{Wolfram}{Wolfram}{1986}]{Wolfram1986}
{\sc Wolfram, S.} (1986).
\newblock {\em Theory and Application of Cellular Automata}.
\newblock World Scientific.

\bibitem[\protect\citeauthoryear{Wolfram}{Wolfram}{2002}]{Wolfram:2002}
{\sc Wolfram, S.} (2002).
\newblock {\em A New Kind of Science}.
\newblock Wolfram Media.
\newblock URL \url{http://www.wolframscience.com/thebook.html}.

\bibitem[\protect\citeauthoryear{Wuensche and Lesser}{Wuensche and
  Lesser}{1992}]{WuenscheLesser1992}
{\sc Wuensche, A.} {\sc and} {\sc Lesser, M.} (1992).
\newblock {\em The Global Dynamics of Cellular Automata; An Atlas of Basin of
  Attraction Fields of One-Dimensional Cellular Automata}.
\newblock Santa Fe Institute Studies in the Sciences of Complexity.
  Addison-Wesley, Reading, MA.

\bibitem[\protect\citeauthoryear{Yukawa, Kikuchi, and Tadaki}{Yukawa
  et~al\mbox{.}}{1994}]{Yukawa:1994}
{\sc Yukawa, S.}, {\sc Kikuchi, M.}, {\sc and} {\sc Tadaki, S.} (1994).
\newblock Dynamical phase transition in one dimensional traffic flow model with
  blockage.
\newblock {\em Journal of the Physical Society of Japan\/}~{\bf 63\/}~(10):
  3609--3618.
\newblock URL \url{http://jpsj.ipap.jp/link?JPSJ/63/3609/}.

\end{thebibliography}

\begin{appendix}

\section*{Appendix}

\section{\emph{Self-organizing} algorithm}
\label{app:sola}

Algorithm \ref{alg:SOLA} formally describes our \emph{self-organizing} method.
Each intersection uses this algorithm independently to regulate traffic, i.e.\ there is no direct communication between intersections. The values of the parameters used in the simulations are shown in Table \ref{table:parameters}. We performed several simulations varying these parameters, and they are quite robust, i.e.\ the performance of the system is not affected by small changes in the parameters.

\IncMargin{0.5cm}
\RestyleAlgo{boxed}
\LinesNumbered
\begin{algorithm}[h!]
\begin{footnotesize}
\Indp

\ForEach{$(\Delta t)$}{
	$t_{i}+= \Delta t$ \tcp*{local phase}

$k_{ij}$ += $vehicles_{approachingRed}$ in $d_j$\tcp*{for rules 1 and 4}

\If {($vehicles_{stoppedAfterGreen}$ at $e > 0$) 
}{
      \If {($vehicles_{stoppedAfterRed}$ at $e_j > 0$, $\forall j$) 	}{
      	$switchAllRed_{i}()$	\tcp*{rule 6}
      }
      \Else{      
      	$switchlight_{i}(j)$ for $j$ with $max(k_{ij})$ \\
	  \hspace{1cm}  and not $vehicles_{stoppedAfterRed}$ at $e_j > 0$	\tcp*{rule 5}
	}
}

\ElseIf{$vehicles_{stoppedAfterRed}$ at $e == 0$}{ \tcp*{no blockage upstream}
	
	  \If{allRed?}{
		$restoreSingleGreen_{i}(j)$ for $j$ with $max(k_{ij})$ \\
	  \hspace{1cm}  and not $vehicles_{stoppedAfterRed}$ at $e_j > 0$
			\tcp*{complement to rule 6}
	  }	

	\If {($k_{ij} \geq 1$) \textbf{and} ($vehicles_{approachingGreen}$ in $d == 0$)  }{
	      $switchlight_{i}(j)$	\tcp*{rule 4}	

	}

	\ElseIf {\textbf{not} ($0 < vehicles_{approachingGreen}$ in $r < m $)}{
		\tcp*{rule 3}
	 \If {($t _{i}\geq t_{\min}$) }{
		\tcp*{rule 2}
	    \If {($k_{ij}$ $\geq $ $n $)}{
	      $switchlight_{i}(j)$ 	\tcp*{rule 1}
	    }
	   }
	  }
	}  
}

$switchlight_{i}(j)$ \Begin{
	$k_{ij} = 0$;	\tcp*{Switch light for direction $j$}
	
	$t_{i} = 0$;
	
	 $switchTrafficLight_{i}(j)$;
	 	
}

  \caption{\emph{Self-organizing} method.}
\label{alg:SOLA}
\end{footnotesize}
\end{algorithm}
\DecMargin{0.5cm}

On every tick ($\Delta t$), Algorithm \ref{alg:SOLA} increases the phase $t_{i}$  by the duration of $\Delta t$ (line 2), and the counters $k_{ij}$ are increased by the number of vehicles approaching or waiting behind a red light $j$ within a certain distance $d_j$ (line 3). \emph{Rule 6} switches all lights to red if every street is blocked ahead of the intersection (line 6). \emph{Rule 5} changes a green light to red if there are vehicles stopped ahead of green light at a distance $e$ from the intersection (lines 9--10). This prevents the accumulation of vehicles when they cannot advance, diminishing the probability of their blocking the intersection, while at the same time allowing vehicles in the crossing street (if any) to advance. The red light with the highest $k_{ij}$ and no blockage upstream is switched to green.
Notice that directions with a green light have no subindex $j$, since there can be at most one direction with a green light, while there can be several directions with red lights.
Rules 5 and 6 are normally used at high vehicle densities. A single green light is restored if there are no vehicles stopped ahead of the intersection and all lights are red (lines 15--16). Again, the light with the highest $k_{ij}$ and no blockage upstream is switched to green.
All of the following rules, to determine whether the light will switch, also depend on the condition that there is free space ahead of the red light.
With \emph{rule 4}, if there are no vehicles approaching a green light within a distance $d$, and there is at least one vehicle approaching a red light ($k_{ij} \geq 1$), this is switched, so that by the time the vehicle(s) reach the intersection it(they) will not need to stop (line 19). This rule is normally used for low vehicle densities. 
\emph{Rule 3} prevents the ``tails" of platoons from being cut, by delaying the switching of a green light when there are few vehicles (fewer than $m$) just about to cross, i.e.\ within a distance $r$ (line 21). Still, rule 3 allows the division of long platoons, preventing the accumulation of vehicles waiting behind a red light. \emph{Rule 2} prevents the fast switching of traffic lights caused by high vehicle densities with a minimum phase $t_{\min}$ (line 22). If rules 2 and 3 are satisfied, \emph{rule 1} changes a traffic light to direction $j$ when the count $k_{ij}$ reaches a certain threshold $n$ (lines 23--24). This makes single vehicles wait for some time, increasing the probability that more vehicles will join them and thus promoting the formation of platoons. Once platoons reach a certain size, they can request a green light before reaching the intersection, if all other conditions are met. Even if this does not occur, once the conditions are proper, the vehicles will get a green light. Thus, in principle vehicles have to wait very little time because of red lights. When a traffic light is switched (lines 30--34), the counter of the direction that will get the green light $k_{ij}$ and the phase $t_{i}$ are reset (lines 31--32). The phase $t_{i}$ keeps the time since the last light switch, independently of the direction selected. Afterwards, the traffic lights are changed (line 33).

\begin{table}[htdp]
\begin{center}
\begin{tabular}{cccc}
\hline
	\textbf{Variable}&	\textbf{Abstract Value}&	\textbf{Scaled Value}&	\textbf{Used by}
\tabularnewline \hline
$\Delta t$	&1 tick		&1/3 s		&		Algorithm
\tabularnewline
$n$	&40 vehicles$\cdot$tick		&13.33 vehicles$\cdot$s		&	Rule 1	
\tabularnewline
$d$	&10 cells		&50 m		&Rules 1 and 4		
\tabularnewline
$t_{\min}$	&10 ticks		&3.33 s		&Rule 2		
\tabularnewline
$m $		&2 vehicles	&2 vehicles		&Rule 3		
\tabularnewline
$r$	&5 cells		&25 m		&Rule 3		
\tabularnewline
$e$	&2 cells		&10 m		&Rules 5	and 6	
\tabularnewline
\hline
\end{tabular}

\caption{Parameters used by \emph{self-organizing} method in simulations.}
\label{table:parameters}
\end{center}
\end{table}

\end{appendix}

\end{document}